\documentclass[%
reprint,
 amsmath,amssymb,
 aps,
prb,
]{revtex4-2}

\usepackage{graphicx}
\usepackage{dcolumn}
\usepackage{bm}
\usepackage{mathrsfs}
\usepackage{braket}
\usepackage{color}
\usepackage{comment}
\usepackage{siunitx}

\usepackage{hyperref}
\hypersetup{
     colorlinks   = true,
     linkcolor    = blue,
     citecolor    = blue,
     urlcolor     = blue
}

\begin{document}

\preprint{APS/123-QED}

\title{Orbital optical activity in noncentrosymmetric metals and superconductors}

\author{Koki Shinada}
\altaffiliation[shinada.koki.64w@st.kyoto-u.ac.jp]{}
\author{Robert Peters}
\affiliation{%
Department of Physics, Kyoto University, Kyoto 606-8502, Japan}%

\date{\today}

\begin{abstract}
We present the optical activity induced by the orbital magnetic moment in metals and superconductors using Green's function formalization. Usually, an apparent singularity of the optical activity vanishes in the normal state; however, we show that it remains finite in the superconducting state and is related to the superconducting Edelstein effect, ensuring the missing area measurement. Finally, we calculate the optical activity in a model Hamiltonian mimicking doped transition metal dichalcogenides to investigate its characteristic spectrum, which reveals a peak shift caused by the interband Fermi surface, its relation with the orbital magnetic moment, and the change of the spectrum in a superconductor at low frequency. We also analyze the optical rotation to discuss the possibility of observing the optical activity in experiments.
\end{abstract}

\maketitle

\section{Introduction}
Optical activity is a ubiquitous optical phenomenon in molecular systems and solids without an inversion center \cite{barron_2004,landau2013electrodynamics}. It originates from the nontrivial coupling between magnetism and electricity and is related to various phenomena, such as optical rotation, dichroism, and birefringence. The optical activity can be roughly divided into two phenomena: natural optical activity (NOA) and the optical magnetoelectric effect (OMEE). 
The NOA includes circular dichroism, which is, e.g., used to discriminate enantiomers of chiral molecules.
The OMEE appears in systems without time-reversal symmetry. Since earlier observations in noncentrosymmetric molecules and solids under the magnetic field or magnetism \cite{pisarev1991optical,krichevtsov1993spontaneous,rikken1997observation}, the OMEE has been observed in a variety of systems \cite{Arima_2008}.

Optical activity is also gaining recognition for its functionality and significance in condensed matter physics. In multiferroics \cite{fiebig2005revival}, spin order accompanying an electric polarization and chirality is spontaneously formed. 
The spin order generates correlations between the magnetism and the electricity, and the OMEE can be finite. 
In multiferroic magnets, the OMEE is also induced by spin waves, called electromagnons, and it forms a resonant peak in the THz regime \cite{pimenov2006possible,PhysRevLett.98.027202,PhysRevLett.98.027203,PhysRevB.80.220406,PhysRevLett.106.057403,takahashi2012magnetoelectric}. In addition, the OMEE is used for visualizing domains of such magnets, which are often difficult to observe with conventional methods due to the absence of macroscopic magnetization \cite{kimura2020imaging,PhysRevB.105.094417,PhysRevResearch.4.043063,PhysRevLett.131.236702}.

Optical activity in solids is not as well developed as in molecular chemistry. However, various theoretical studies have been conducted in recent years. 
The band theory of optical activity is discussed in several works \cite{doi:10.1143/JPSJ.39.1013,PhysRevB.48.1384,PhysRevB.81.094525,PhysRevB.82.245118,PhysRevB.88.134514,PhysRevLett.116.077201,PhysRevB.92.235205,PhysRevLett.122.227402,PhysRevB.106.085413,10.21468/SciPostPhys.14.5.118}, and it has been applied to a variety of systems, including chiral crystals \cite{PhysRevB.79.075438,PhysRevB.97.035158,rerat2021first,PhysRevB.107.045201}, twisted bilayer graphene \cite{morell2017twisting,PhysRevLett.120.046801,PhysRevB.106.245405,PhysRevB.107.195141}, and topological antiferromagnets \cite{ahn2022theory}. 
Recently, the formalization has been extended to metals \cite{PhysRevB.106.245405,10.21468/SciPostPhys.14.5.118}.
In addition, it is becoming clear that optical activity also plays an essential role in exotic phases, including topological materials and superconductors. For example, it is revealed that the circular dichroism in chiral multifold semimetals is quantized and that it can be used to probe the topological nature \cite{PhysRevB.105.235403,mandal2023signatures,PhysRevLett.131.116603}. Furthermore, for superconductors, it has recently been shown that the optical activity can be applied to detect the superconducting Edelstein effect, which is unique to noncentrosymmetric superconductors and has not been observed experimentally, using the missing area measurement \cite{PhysRevB.108.165119}. Despite such progress regarding optical activity, research in crystals is less advanced than in molecular chemistry, and further development is required.

In this paper, we discuss the optical activity originating from orbital magnetic moments in metals and superconductors using a Green's function formalism. We discuss a no-go theorem concerning a singularity of the optical activity in Secs.~\ref{oa_normal_sec} and \ref{oa_sc_sec}, and we show that the singularity appearing in the superconducting state is related to the superconducting Edelstein effect.
Furthermore, we present model calculations investigating two-dimensional materials such as transition metal dichalcogenides and twisted bilayer graphene in the normal state and the superconducting state in Secs.~\ref{model_normal} and \ref{model_sc}, respectively. We also analyze the spectrum of the optical rotation. Finally, in Sec.~\ref{conclusion}, we conclude this paper and discuss the possibility of an experimental observation.

\section{orbital optical activity in the normal state} \label{oa_normal_sec}
In this section, we derive the orbital optical activity in the normal state using a Green's function method. The optical activity originates from the wavenumber dependence of the optical conductivity $\sigma(\bm{q},\Omega)$, where $\bm{q}$ and $\Omega$ are the wavenumber and the frequency of the irradiation light. $\sigma(\bm{q}=0,\Omega)$ represents the uniform conductivity that changes the refractive index of a material, leading to observable effects such as the reflectance and the rotation of linearly-polarized light in magnets.
In noncentrosymmetric systems, the first-order term in the wavenumber of the optical conductivity $\bm{\partial}_{q} \sigma(\bm{q}=0,\Omega) \cdot \bm{q}$ can be significant. This is the optical activity, which is the main quantity in this paper, exhibiting optical rotation, dichroism, and birefringence. It includes two parts: a spin part and an orbital part. In this paper, we only focus on the orbital part.

\subsection{Green's function formalism of the orbital optical activity}
We consider noninteracting systems. The Hamiltonian without an electromagnetic field is assumed to be described by the nonrelativistic quadratic kinetic energy term with a spin-orbit coupling term as
\begin{equation}
    H_0 = \frac{\bm{p}^2}{2m} + V(\bm{x}) + \frac{1}{4m^2} \biggl( \frac{\partial V(\bm{x})}{\partial \bm{x}} \times \bm{p} \biggr) \cdot \bm{\sigma}.
\end{equation}
Here, $m$ is the electron mass, $V(\bm{x})$ is a periodic potential, and $\bm{\sigma}$ are the Pauli matrices representing the spin degrees of freedom. This periodic Hamiltonian is diagonalized by the Bloch theorem using the Bloch wave function $\ket{\psi_{n\bm{k}}}$ as $H_0 \ket{\psi_{n\bm{k}}} = \epsilon_{n\bm{k}} \ket{\psi_{n\bm{k}}}$, where $n$ is the band index and $\bm{k}$ is the Bloch wave number. The Bloch wave function is split into the product of a plane wave and a periodic function $\ket{u_{n\bm{k}}}$ as $\ket{\psi_{n\bm{k}}} = e^{i\bm{k} \cdot \bm{x}} \ket{u_{n\bm{k}}}$. We define the Bloch Hamiltonian $H_{\bm{k}} = e^{-i\bm{k} \cdot \bm{x}} H_0 e^{i\bm{k} \cdot \bm{x}}$, which is diagonalized by $\ket{u_{n\bm{k}}}$ as $H_{\bm{k}} \ket{u_{n\bm{k}}} = \epsilon_{n\bm{k}} \ket{u_{n\bm{k}}}$.

We introduce the electromagnetic field using the vector potential $\bm{A}(\bm{x},t)$ by shifting the momentum $\bm{p} \to \bm{p} + e\bm{A}(\bm{x},t)$ ($-e<0$ is the electron charge). We use the temporal gauge, where the vector potential alone fully describes the electromagnetic field without a scalar potential. The vector potential yields an additional term up to the first order
\begin{equation}
    H_{A} = \frac{1}{2} (\bm{v} \cdot \bm{A}(\bm{x},t) + \bm{A}(\bm{x},t) \cdot \bm{v}),
\end{equation}
where $\bm{v} = i [H_0 , \bm{x}]$ is the velocity operator. Here, we ignore the diamagnetic term, which is the quadratic term, because it only contributes to the conductivity independent of $\bm{q}$ and, thus, does not contribute to the optical activity.
We also neglect the Zeeman term, which is the coupling between the magnetic field and the spin moment and contributes to the spin part of the optical activity \cite{10.21468/SciPostPhys.14.5.118,PhysRevB.108.165119}. 
Focusing only on the orbital part of the optical activity in this paper, we can neglect this term.

The current operator conjugate with the vector potential $\bm{A}(\bm{r},t)$ ($\bm{r}$ is just a position coordinate, not an operator) is given by 
\begin{equation}
    \bm{J}(\bm{r}) \equiv - \frac{\delta H_A}{\delta \bm{A}(\bm{r},t)} = -\frac{e}{2} \{ \bm{v} , \delta(\bm{r} - \bm{x}) \}.
\end{equation}
Using the dynamical linear response theory, the current-current correlation function reads
\begin{align}
    &\Phi_{\mu \nu}(\bm{q},\Omega) \notag \\
    &= e^2 \int [d^4k] f(\omega) \mathrm{Tr} \Bigl[
    G^{RA}(\bm{k}-,\omega) v^{\mu}_{\bm{k}} G^R(\bm{k}+,\omega+\Omega) v^{\nu}_{\bm{k}} \notag \\
    &\quad +G^A(\bm{k}-,\omega-\Omega) v^{\mu}_{\bm{k}} G^{RA}(\bm{k}+,\omega) v^{\nu}_{\bm{k}}
    \Bigr].
\end{align}
Here, $G^{R/A}(\bm{k},\omega)=1/(\omega - H_{\bm{k}} + \mu \pm i \Gamma)$ is the retarded/advanced Green's function at a chemical potential $\mu$, and $\bm{v}_{\bm{k}} = \partial H_{\bm{k}}/\partial \bm{k}$ is the velocity operator of the Bloch Hamiltonian. $f(\omega) = 1/(e^{\beta \omega} +1)$ is the Fermi distribution function at temperature $1/\beta$. We use the following abbreviations; $\int [d^4k] = \int^{\infty}_{-\infty} d\omega /(2\pi i) \int_{\mathrm{BZ}} d^3 k /(2\pi)^3$, $G^{RA} = G^R -G^A$, and $\bm{k} \pm = \bm{k} \pm \bm{q}/2$. In this paper, we phenomenologically introduce the dissipation effect by a finite $\Gamma$.

Expanding the correlation function by the wavenumber $\bm{q}$, the first-order term is given by
\begin{align}
    \Phi_{\mu \nu \lambda}(\Omega)
    &= \frac{e^2}{2} \int [d^4k] f(\omega)
    \nonumber \\
    &\times \mathrm{Tr}
    \Bigl[
    -\partial_{\lambda}G^{RA}(\bm{k},\omega) v_{\bm{k}}^{\mu} G^{R}(\bm{k},\omega+\Omega) v_{\bm{k}}^{\nu} 
    \nonumber \\
    &+G^{RA}(\bm{k},\omega) v_{\bm{k}}^{\mu} \partial_{\lambda} G^R(\bm{k},\omega+\Omega) v_{\bm{k}}^{\nu}
    \nonumber \\
    &-\partial_{\lambda} G^A(\bm{k},\omega-\Omega) v_{\bm{k}}^{\mu} G^{RA}(\bm{k},\omega) v^{\nu}_{\bm{k}}
    \nonumber \\
    &+G^A(\bm{k},\omega-\Omega) v_{\bm{k}}^{\mu} \partial_{\lambda} G^{RA}(\bm{k},\omega) v_{\bm{k}}^{\nu}  
    \Bigr].
\end{align}
Here, we use $\partial_{\lambda} = \partial/\partial k_{\lambda}$. We note the identity $\partial_{\lambda} G^{R/A} = G^{R/A} v_{\bm{k}}^{\lambda} G^{R/A}$, which is useful for numerical calculations. Finally, we obtain the optical activity $\sigma_{\mu \nu \lambda}(\Omega) = \Phi_{\mu \nu \lambda}(\Omega)/i(\Omega+ i\delta)$, where $\delta = +0$ is an adiabatic factor. This tensor $\sigma_{\mu \nu \lambda}$ describes all optical activity phenomena induced by the orbital moment, including the NOA (antisymmetric part by interchange of indices $\mu$ and $\nu$) and the OMEE (symmetric part).
This formulation has been useful for analyses in periodic crystals as it avoids difficulties such as origin dependence and gauge invariance.
Some works have derived this tensor with the band representation and proved explicitly that this formalism avoids these difficulties, providing a physical interpretation of the obtained gauge-invariant equations \cite{PhysRevB.82.245118,PhysRevLett.116.077201,PhysRevB.107.094106,PhysRevB.107.214109,10.21468/SciPostPhys.14.5.118}.

\subsection{No-go theorem} \label{nogo_sec}
The optical activity has a singularity at $\Omega = 0$ because of the factor $1/(\Omega + i\delta) = 1/\Omega - i\pi \delta(\Omega)$. This $\delta$-function and the accompanying anomalous divergence scaling $\Omega^{-1}$ are known as the false divergences problem when using the velocity gauge from the minimal-coupling Hamiltonian \cite{PhysRevB.106.245405}. 
This divergence is supposed to vanish in the normal state (no-go theorem), which has been shown using the Bloch band representation \cite{10.21468/SciPostPhys.14.5.118}. 
We can also reconfirm that the singularity vanishes for almost all of the components of $\sigma_{\mu \nu \lambda}$ using the Green's function formula as seen in the following.

The coefficient in front of the $\delta$ function in the optical activity reads
\begin{equation}
    \Phi_{\mu \nu \lambda}(0) = i \varepsilon_{\mu \nu \lambda} \alpha.
\end{equation}
Here, $\varepsilon_{\mu \nu \lambda}$ is the totally antisymmetric tensor. Almost all components of $\Phi_{\mu \nu \lambda}(0)$ are zero.
Only totally antisymmetric terms appear due to $\alpha$, which is real (see Appendix \ref{nogo} for a detailed derivation).
This term corresponds to the chiral magnetic effect, where a current density is induced by a magnetic field as $\bm{J} = \alpha \bm{B}$. However, this phenomenon usually does not occur unless, for example, the chiral chemical potential $\mu_5$ is finite in the non-equilibrium case \cite{PhysRevD.78.074033}. Using the Bloch band representation, this $\alpha$ is shown to be zero, resulting from its topological nature \cite{PhysRevLett.116.077201,10.21468/SciPostPhys.14.5.118}. However, the formula using Green's function can not immediately prove that this term vanishes. 
In conclusion, the no-go theorem can be checked for all components except for the totally-antisymmetric term.
We leave the general proof whether this term vanishes using Green's functions assuming, e.g., strongly correlated systems as a future work. Thus, in the following discussion, we only consider cases where the totally-antisymmetric component vanishes due to crystal point group symmetry.

\section{orbital optical activity in a superconductor} \label{oa_sc_sec}
\subsection{Green's function formula for a superconductor}
In this paper, we treat the superconducting state using the mean-field approximation represented by the Bogoliubov-de Gennes (BdG) Hamiltonian as
\begin{subequations}
\begin{align}
    H_{\mathrm{BdG}} &= \frac{1}{2} \sum_{\bm{k}nm} \psi^{\dagger}_{n\bm{k}} H^{\mathrm{BdG}}_{\bm{k}nm} \psi_{m\bm{k}}, \\
    H^{\mathrm{BdG}}_{\bm{k}} &=
    \begin{pmatrix}
        H_{\bm{k}} - \mu & -\Delta_{\bm{k}} \\
        -\Delta^{\dagger}_{\bm{k}} & -H^{\mathrm{T}}_{-\bm{k}} + \mu
    \end{pmatrix}.
\end{align}
\end{subequations}
Here, $\bm{\psi}^{\dagger}_{\bm{k}} = (c_{1 \bm{k}}^{\dagger}, \cdots c_{N\bm{k}}^{\dagger}, c_{1-\bm{k}}, \cdots , c_{N-\bm{k}})$ is the Nambu spinor and $\Delta_{\bm{k}}$ is the pair potential, which is the order parameter of the superconducting state. We define $M^{\mathrm{T}}$ and $M^{\dagger}$ as the transpose and the Hermitian conjugate of a matrix $M$. The current-current correlation function in the superconducting state is given by
\begin{align}
        &\Phi_{\mu \nu}(\bm{q},\Omega) \notag \\
        &=\frac{e^2}{2} \int [d^4k] f(\omega) \mathrm{Tr}
        \Bigl[
        G^{RA}_{\mathrm{BdG}}(\bm{k}-,\omega) \tilde{v}^{\mu}_{\bm{k}} G^{R}_{\mathrm{BdG}}(\bm{k}+,\omega+\Omega) \tilde{v}^{\nu}_{\bm{k}} \notag \\
        & \quad + G^A_{\mathrm{BdG}}(\bm{k}-,\omega-\Omega) \tilde{v}^{\mu}_{\bm{k}} G^{RA}_{\mathrm{BdG}}(\bm{k}+,\omega) \tilde{v}^{\nu}_{\bm{k}}
        \Bigr].
\end{align}
Here, $G^{R/A}_{\mathrm{BdG}}(\bm{k},\omega) = 1/(\omega - H^{\mathrm{BdG}}_{\bm{k}} + \Sigma^{R/A}(\bm{k},\omega))$ is the retarded/advanced Green's function for the BdG Hamiltonian, including the self-energy describing the dissipation effect. A specific equation of this self-energy will be given in a model calculation below. The velocity operator is different from the normal state. It is given by
\begin{equation}
    \tilde{v}^{\mu}_{\bm{k}} =
    \begin{pmatrix}
        v^{\mu}_{\bm{k}} & 0 \\
        0 & - (v^{\mu}_{-\bm{k}})^{\mathrm{T}}
    \end{pmatrix}.
\end{equation}
We note that this velocity operator is not the usual derivative of the BdG Hamiltonian $H^{\mathrm{BdG}}_{\bm{k}}$ by the wavenumber $\bm{k}$. Expanding the correlation function by $\bm{q}$, the first-order term is 
\begin{align}
    \Phi_{\mu \nu \lambda}(\Omega) &=
    \frac{e^2}{4} \int [d^4k] f(\omega)
    \nonumber \\
    &\times\mathrm{Tr} \Bigl[
    -\partial_{\lambda} G^{RA}_{\mathrm{BdG}}(\bm{k},\omega) \tilde{v}^{\mu}_{\bm{k}} G^{R}_{\mathrm{BdG}}(\bm{k},\omega+\Omega) \tilde{v}^{\nu}_{\bm{k}}
    \nonumber \\
    &+G^{RA}_{\mathrm{BdG}}(\bm{k},\omega) \tilde{v}^{\mu}_{\bm{k}} \partial_{\lambda} G^R_{\mathrm{BdG}}(\bm{k},\omega+\Omega) \tilde{v}^{\nu}_{\bm{k}}
    \nonumber \\
    &-\partial_{\lambda} G^A_{\mathrm{BdG}}(\bm{k},\omega-\Omega) \tilde{v}^{\mu}_{\bm{k}} G^{RA}_{\mathrm{BdG}}(\bm{k},\omega) \tilde{v}^{\nu}_{\bm{k}}
    \nonumber \\
    &+G^{A}_{\mathrm{BdG}}(\bm{k},\omega -\Omega) \tilde{v}^{\mu}_{\bm{k}} \partial_{\lambda} G^{RA}_{\mathrm{BdG}}(\bm{k},\omega) \tilde{v}^{\nu}_{\bm{k}}
    \Bigr]. \label{oa_green_sc}
\end{align}
The derivative of the Green's function by the wavenumber can be transformed into the useful form for numerical calculations as $\partial_{\lambda} G^{R/A}_{\mathrm{BdG}} = G^{R/A}_{\mathrm{BdG}} \hat{v}_{\bm{k}}^{\lambda} G^{R/A}_{\mathrm{BdG}}$, where
\begin{equation}
    \hat{v}^{\lambda}_{\bm{k}} = \frac{\partial H^{\mathrm{BdG}}_{\bm{k}} }{\partial k_{\lambda}} = 
    \begin{pmatrix}
        v^{\lambda}_{\bm{k}} & - \partial_{\lambda} \Delta_{\bm{k}} \\
        - \partial_{\lambda} \Delta_{\bm{k}}^{\dagger} & (v^{\lambda}_{-\bm{k}})^{\mathrm{T}}
    \end{pmatrix}.
\end{equation}
This velocity operator $\hat{v}$ is different from $\tilde{v}$ even if the pair potential is independent of the wavenumber. Finally, we obtain the optical activity for the superconducting state by $\sigma_{\mu \nu \lambda}(\Omega) = \Phi_{\mu \nu \lambda}(\Omega)/i(\Omega + i \delta)$.

\subsection{Singularity and relation to the superconducting Edelstein effect} \label{SEE}
In the previous section, we have seen that the optical activity appears to have a $\delta$-function singularity at $\Omega=0$, which, however, vanishes in the normal state. 
In the superconducting case, the singularity generally is finite.
The absence of cancellation is attributed to the difference in the velocity operator from the normal state.
The coefficient of the $\delta$ function satisfies (see Appendix \ref{nogo} for a detailed derivation)
\begin{equation}
    \Phi_{\mu\nu\lambda}(0) = \tilde{\Phi}_{\mu\nu\lambda} - (\mu \leftrightarrow \nu).
\end{equation}
$\tilde{\Phi}_{\mu \nu \lambda}$ is purely imaginary, and its detailed form is discussed in Appendix \ref{nogo}.
$\Phi_{\mu \nu \lambda}(0)$ is antisymmetric for the interchange between $\mu$ and $\nu$.
Being purely imaginary is consistent with the general symmetry argument for a current-current correlation function
\cite{PhysRevB.108.165119}.
Therefore, the $\delta$-function singularity appears in the imaginary part of the optical activity $\mathrm{Im} \sigma_{\mu \nu \lambda} (\Omega)$, and the anomalous $1/\Omega$ divergence with the same coefficient $\Phi_{\mu \nu \lambda}(0)$ as the $\delta$-function singularity appears in the real part $\mathrm{Re} \sigma_{\mu\nu\lambda}(\Omega)$  to satisfy the Kramers-Kronig relation.
Due to the antisymmetry in the indices $\mu$ and $\nu$, the anomalous term only appears in the NOA. The OMEE does not exhibit this singularity because the OMEE originates from the symmetric part of the interchange between $\mu$ and $\nu$ due to the necessity of the time-reversal symmetry breaking.
Such anomalous frequency dependence is common in superconductors. The optical conductivity has the same anomaly, and the coefficient corresponds to the superfluid density \cite{tinkham2004superconductivity}. Furthermore, it has recently been shown that also the nonlinear conductivity has a similar anomaly, and the coefficient has the meaning of the nonreciprocal superfluid density \cite{PhysRevB.105.024308,PhysRevB.105.L100504}.
The physical meaning of the coefficient in the case of the optical activity is also known. It has been studied for the spin contribution, and the coefficient corresponds to the superconducting Edelstein effect \cite{PhysRevB.81.094525,PhysRevB.88.134514,PhysRevB.108.165119}. In this paper, we consider the orbital contribution, and we will reveal the physical meaning of $\Phi_{\mu \nu \lambda}(0)$ in the following discussion.

$\Phi_{\mu \nu \lambda}(0)$ has nine components due to its antisymmetric property. 
Therefore, it can be rewritten by a rank-2 tensor $\mathcal{K}_{\mu \nu}$ with a one-to-one correspondence \cite{PhysRev.171.1065,PhysRevB.82.245118} as
\begin{subequations}
\begin{align}
    &\Phi_{\mu \nu \lambda}(0) = i \varepsilon_{\mu \lambda \theta} \mathcal{K}_{\nu \theta} - i \varepsilon_{\nu \lambda \theta} \mathcal{K}_{\mu \theta}, \label{decomposition} \\
    &\mathcal{K}_{\mu \nu} = -\frac{1}{4i} \varepsilon_{\nu \theta \lambda} \Bigl( 2 \Phi_{\mu \theta \lambda}(0) - \Phi_{\theta \lambda \mu}(0) \Bigr).
\end{align}
\end{subequations}
Here, $\mathcal{K}_{\mu \nu}$ is real because $\Phi_{\mu \nu \lambda}(0)$ is purely imaginary. Substituting Eq.~(\ref{decomposition}) into the original linear response relation ($J_{\mu}= \Phi_{\mu \nu}(\bm{q},\Omega) A_{\nu}$), we obtain
\begin{equation}
    J_{\mu} = \mathcal{K}_{\mu \theta} B_{\theta} + (i \bm{q} \times \bm{M})_{\mu},\quad M_{\theta} = \mathcal{K}_{\mu \theta} A_{\mu}. 
\end{equation}
This equation shows that $\mathcal{K}$ represents the superconducting Edelstein effect (SEE), where a supercurrent induces a magnetization $\bm{M}$ or the magnetic field $\bm{B}$ induces the supercurrent. 
Thus, $\Phi_{\mu \nu \lambda}(0)$ also has the physical meaning of the SEE for the orbital contribution, in analogy to the case of the spin contribution.

Previous works about the SEE \cite{levitov1985magnetoelectric,PhysRevLett.75.2004,PhysRevLett.87.037004,PhysRevB.65.144508,PhysRevB.72.024515,PhysRevB.77.054515} have mainly focused on the induced spin moment in systems with spin-orbit coupling (SOC). Recently, supercurrent-induced orbital magnetization has been proposed \cite{PhysRevResearch.3.L032012} and is expected to appear in systems where the valley degrees of freedom are of key importance, such as the transition metal dichalcogenides (TMDs) and twisted bilayer graphene (TBG), even without SOC.
The SEE is a unique response for noncentrosymmetric superconductors and is expected to provide essential information about them; however, it has not been observed in experiments for both spin and orbital parts.

Here, we propose another means of observing this SEE. 
By measuring the spectrum of the optical activity, we can obtain $\Phi_{\mu \nu \lambda}(0)$, which is equivalent to the SEE.
$\Phi_{\mu \nu \lambda}(0)$ can be obtained by measuring the coefficient of the anomalous divergence of $\mathrm{Re} \sigma_{\mu \nu \lambda}(\Omega)$ scaling as $\Omega^{-1}$. Furthermore, the spectrum of $\mathrm{Im} \sigma_{\mu \nu \lambda}(\Omega)$ can also determine the value of $\Phi_{\mu\nu\lambda}(0)$. 
Reference \cite{PhysRevB.108.165119} has shown that the missing area, the area reduced from the spectrum of the normal phase created by the superconducting gap, is identical to $\Phi_{\mu \nu \lambda}(0)$ because of a universal sum rule and the no-go theorem in the case of the spin contribution. This method is useful for dirty superconductors, which is a common situation.
As discussed in Sec.~\ref{nogo_sec} and this section, the no-go theorem is also established for the orbital contribution. Thus, the missing area measurement is also valid for the orbital contribution.

\section{model calculation in the normal state} \label{model_normal}
In this section, we discuss the orbital optical activity using a model Hamiltonian to investigate the characteristics of the spectrum. 
\begin{figure*}[t]
\includegraphics[width=0.28\linewidth]{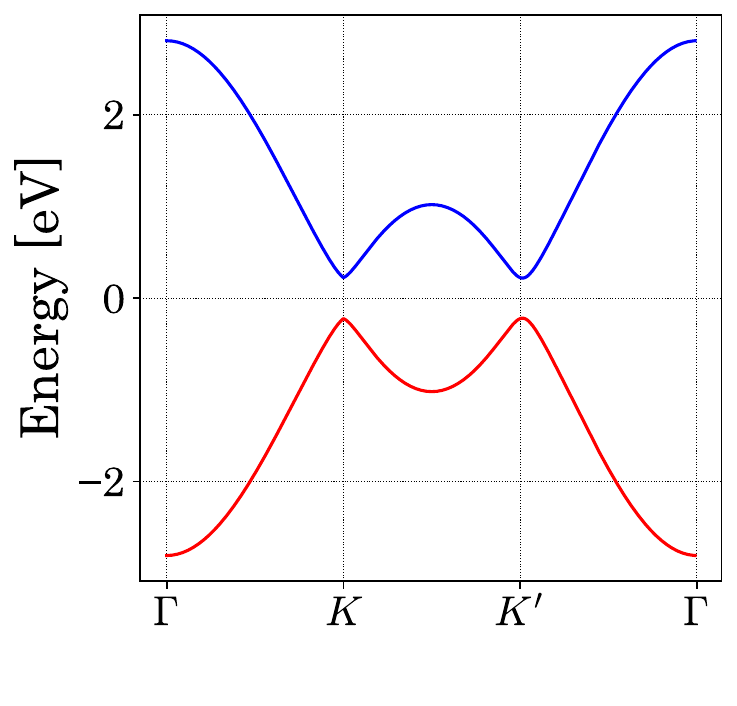}
\includegraphics[width=0.35\linewidth]{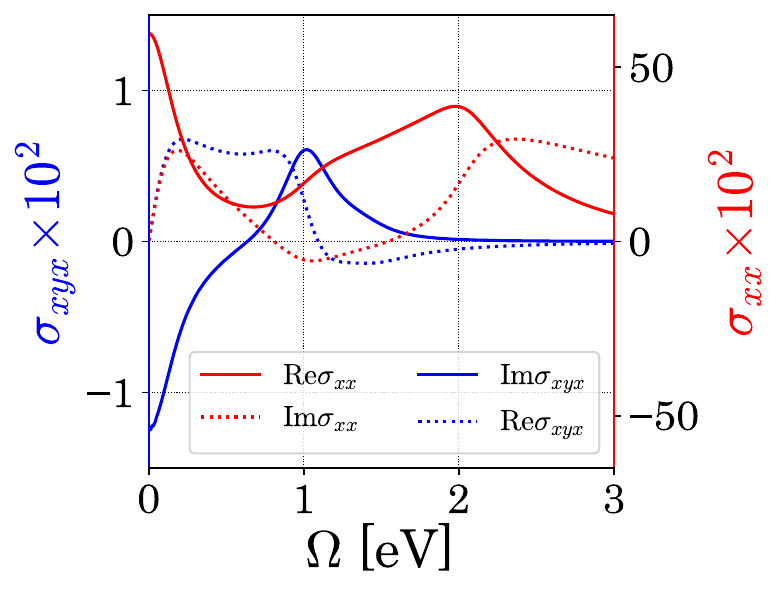}
\includegraphics[width=0.35\linewidth]{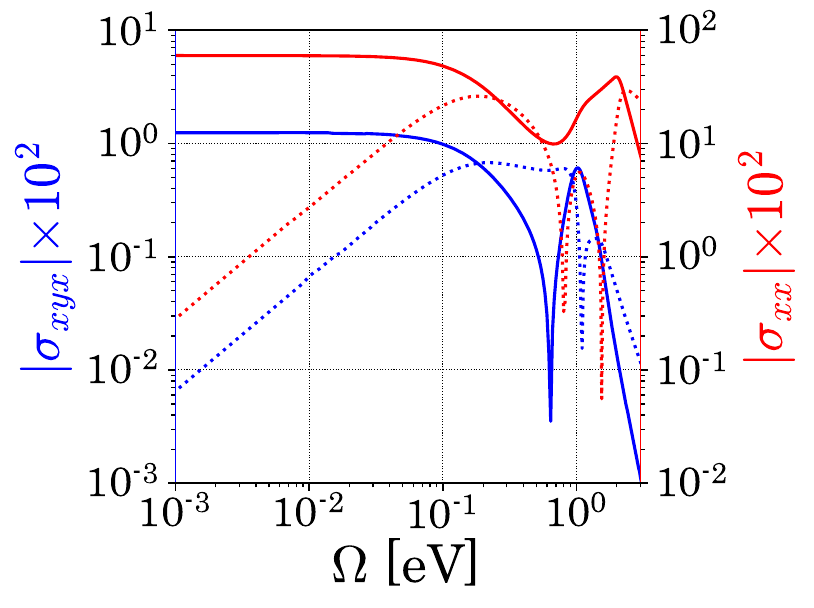}
\caption{(Left) Band structure of the used model: We set $t=1\mathrm{eV}$, $t' = 0.9t$, and $G=0.2t$. (Middle) Numerical results of the orbital optical activity $\sigma_{xyx}$ (blue lines and left axis) and the optical conductivity $\sigma_{xx}$ (red lines and right axis): We set $t=1\mathrm{eV}$, $t' = 0.9t$, $G=0.2t$, $\mu=0.5t$, and $\Gamma=0.1t$. (Right) The magnitude of $\sigma_{xyx}$ and $\sigma_{xx}$ is plotted on a logarithmic scale. The lines are the same as in the middle panel. We set $\sigma_{xyx}$ and $\sigma_{xx}$ in units of $e^2a/\hbar$ and $e^2/\hbar$.
}  \label{oa_normal_fig}
\end{figure*}

\subsection{Optical activity in the normal state}
We discuss the orbital optical activity in the normal state. We use a model Hamiltonian imitating strained $\mathrm{MoS_2}$. For simplicity, we ignore the SOC, which is realized in the conduction band of $\mathrm{MoS_2}$ \cite{PhysRevB.85.205302,KADANTSEV2012909}. In this case, the spin part of the optical activity vanishes, and the orbital part plays the central role.

The Hamiltonian is written as
\begin{subequations}
  \begin{gather}
    H_{\bm{k}} =
    \begin{pmatrix}
        G & \epsilon_{\bm{k}} \\
        \epsilon^*_{\bm{k}} & -G 
    \end{pmatrix} \otimes \sigma_0, \\
    \epsilon_{\bm{k}} = t e^{i\bm{k} \cdot \bm{a}_1} + t' ( e^{i\bm{k} \cdot \bm{a}_2} + e^{i\bm{k} \cdot \bm{a}_3} ).
  \end{gather}
\end{subequations}
Here, $t$ and $t'$ are hopping parameters, and $G$ is the gap parameter that opens a gap at the K and $\mathrm{K}'$ points due to spatial inversion symmetry breaking. We use $\bm{a}_1=(0,1/\sqrt{3})$, $\bm{a}_2 = (-1/2,-1/2\sqrt{3})$ and $\bm{a}_3 = (1/2,-1/2\sqrt{3})$, and we set the lattice constant $a=1$ ($a = 2.4~\mathrm{\AA}$ for $\mathrm{MoS_2}$). $\sigma_0$ is the identity matrix for the spin degrees of freedom. In the following, we set $t > t'$ to incorporate the effect of stretching in the $x$ direction.
We plot the band structure in the left panel of Fig.~\ref{oa_normal_fig}.
The Hamiltonian belongs to the point group $\mathrm{C_{2v}}$. Therefore, the Edelstein effect and the orbital optical activity are restricted to the one component $\chi_{xz}$, where $J_{\mu} = -\chi_{\mu \nu} \dot{B}_{\nu}$ and $M_{\nu} = \chi_{\mu \nu} E_{\mu}$. Then, the nonzero component of the optical activity tensor is only $\sigma_{xyx} = -\sigma_{yxx}$ and is related to $\chi_{xz} = -i \sigma_{xyx}$ \cite{PhysRevResearch.2.012073}. The stretching generates the in-plane polarization and the induced out-of-plane magnetization. This type of optical activity is different from the case of the chiral-structured materials, where the induced magnetization is parallel to the electric field \cite{kim2016chiral,PhysRevB.106.245405}.

We show the numerical results of the optical activity $\sigma_{xyx}$ in the middle and right panels of Fig.~\ref{oa_normal_fig} (blue lines). We use $t=1~\mathrm{eV}$, $t' = 0.9t$, $G=0.2t$, $\mu=0.5t$, and $\Gamma=0.1t$. 
At frequencies smaller than the dissipation, $\Omega < \Gamma$, intraband transitions are dominant and show a Drude model-like behavior. 
In this region, the imaginary part of the optical activity is larger than the real part, and the real part is linear in the frequency and becomes comparable with the imaginary part when the frequency reaches the scale of the dissipation. 
This behavior is identical to that of the uniform optical conductivity described by the Drude model. 
It is consistent with the equation of the gyrotropic magnetic effect (GME) in a nonabsorbing regime \cite{PhysRevLett.116.077201}.
This effect originates from the orbital magnetic moment dipole (OMMD) 
\begin{equation}
\int_{\mathrm{BZ}} \frac{d^d k}{(2\pi)^d} \sum_n f_{n\bm{k}} \partial_{k_i} m^{\mathrm{orb}}_{n\bm{k}},
\end{equation}
where $f_{n\bm{k}} = f(\epsilon_{n\bm{k}}-\mu)$ and $m^{\mathrm{orb}}_{n\bm{k}}$ is the orbital magnetic moment oriented in the direction perpendicular to the surface. 
Without stretching $\mathrm{MoS_2}$, orbital magnetic moments (OMMs) with opposite signs exist around the $\mathrm{K}$ and $\mathrm{K}'$ points \cite{PhysRevLett.99.236809}; however, the OMMD vanishes due to $\mathrm{C_{3v}}$ symmetry.
When stretching $\mathrm{MoS_2}$, the OMMD in the same direction as the stretching direction becomes finite and generates the GME. This discussion is similar to the Berry curvature dipole in the nonlinear Hall effect \cite{PhysRevLett.115.216806}. The GME in the DC limit has been experimentally observed as the orbital Edelstein effect in stretched $\mathrm{MoS_2}$ \cite{PhysRevLett.123.036806,lee2017valley}.
Above the frequency corresponding to the optical gap $\sim 2\mu$, interband effects become finite around $\mathrm{K}$ and $\mathrm{K}'$ points.

For the following discussion, we also calculate the optical conductivity.
In Fig.~\ref{oa_normal_fig} (red lines), we plot the numerical results of the longitudinal optical conductivity $\sigma_{xx}$. The character of this spectrum is similar to the optical activity. In the low-frequency regime, $\Omega < \Gamma$, the optical conductivity follows the Drude theory. 
The magnitude of the conductivity in the DC limit is $0.6 e^2/\hbar$, which is comparable to a dual-gated monolayer $\mathrm{MoS_2}$ \cite{radisavljevic2013mobility}.
Above the frequency corresponding to the optical gap ($\sim 2\mu$), interband transitions appear, and the spectrum forms a Lorentz-peak structure.

However, a clear difference from the optical activity can be identified. We can see that the interband peak position differs between $\mathrm{Im} \sigma_{xyx}$ and $\mathrm{Re} \sigma_{xx}$. The peaks are located at $\Omega = 1~\mathrm{eV}$ and $\Omega = 2~\mathrm{eV}$, respectively. 
\begin{figure}[t]
\includegraphics[width=0.49\linewidth]{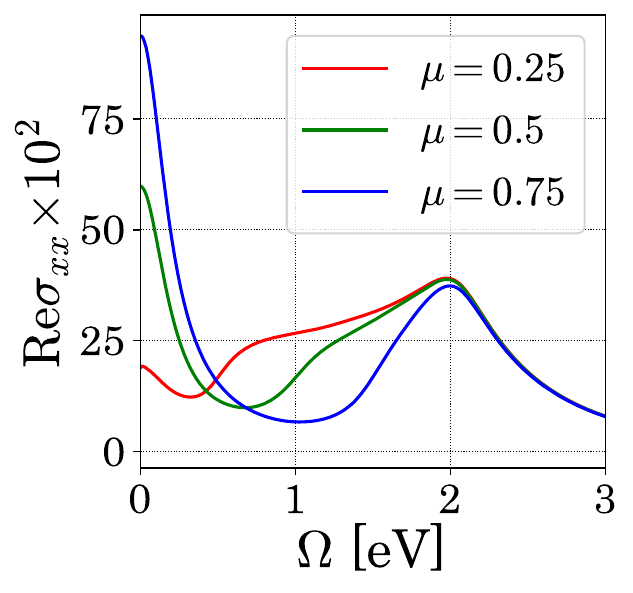}
\includegraphics[width=0.49\linewidth]{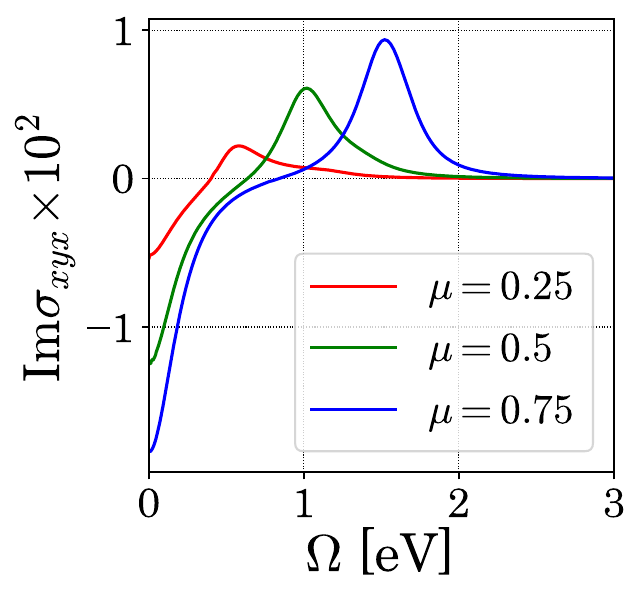}
\caption{Chemical potential dependence of the optical conductivity (Left) and the optical activity (Right). We use the same parameters and units as in Fig.~\ref{oa_normal_fig}.}  \label{mudepend_fig}
\end{figure}
This distinction becomes clearer when looking at the chemical potential dependence in Fig.~\ref{mudepend_fig}. While the interband peak of the optical conductivity $\mathrm{Re}\sigma_{xx}$ does not depend on the chemical potential and is located at $\Omega = 2~\mathrm{eV}$, the interband peak in the optical activity $\mathrm{Im} \sigma_{xyx}$ shifts as the chemical potential is increased. It is located at the optical gap $2\mu$. The typical optical transition occurs as a transition from an occupied state to an unoccupied state. Thus, the spectra should match once the optical gap is exceeded. The spectra of $\mathrm{Re}\sigma_{xx}$ are indeed identical among different chemical potentials above the optical gap $2\mu$; however, the spectra of $\mathrm{Im} \sigma_{xyx}$ behave differently.
In the following, we explain the reason for these distinctions.

\begin{figure*}[t]
\includegraphics[width=0.29\linewidth]{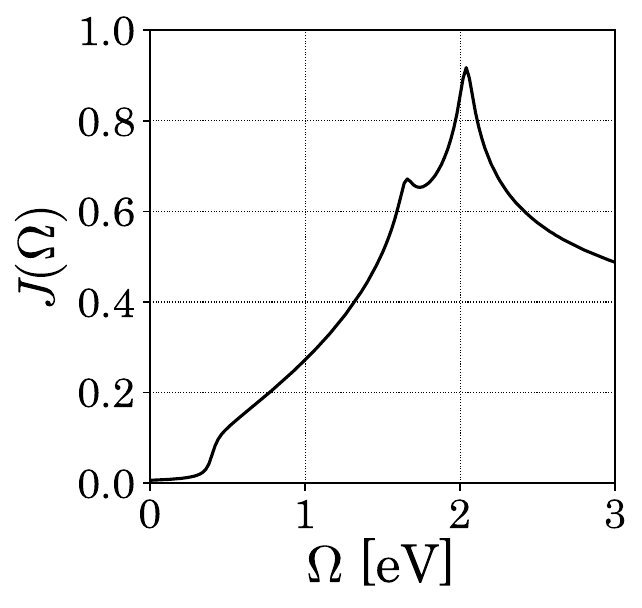}
\includegraphics[width=0.33\linewidth]{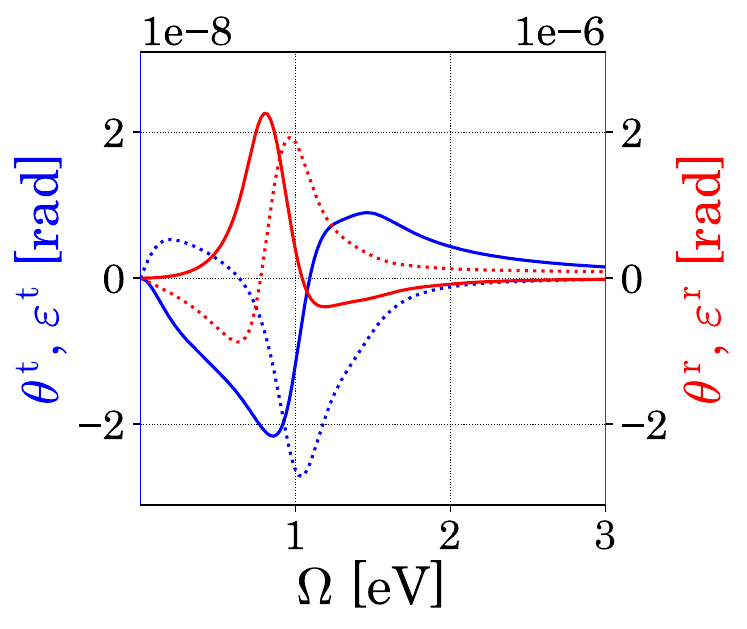}
\includegraphics[width=0.3\linewidth]{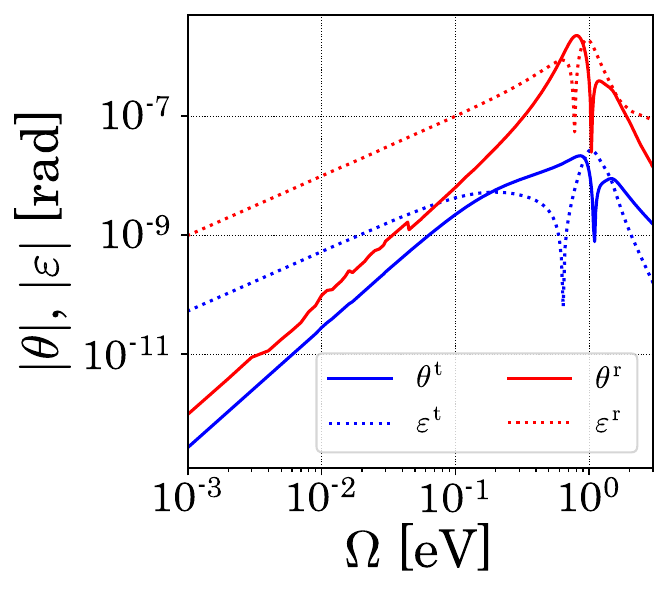}
\caption{(Left) Numerical results of the JDOS, $J(\Omega)$. We use the same parameters as in Fig.~\ref{oa_normal_fig}. (Middle) Numerical results of the optical rotational angle $\theta$ and the ellipticity $\varepsilon$. The red and blue lines correspond to the reflected light (right axis) and the transmitted light (left axis). The legend is identical to the right panel. We use the same parameters as in Fig.~\ref{oa_normal_fig}. (Right) The magnitude of $\theta$ and $\varepsilon$ is plotted on a logarithmic scale.}  \label{jdos_fig}
\end{figure*}

The interband transition peak is often explained by the joint density of states (JDOS). The JDOS is defined by
\begin{equation}
    J(\Omega) = \int_{\mathrm{BZ}} \frac{d^dk}{(2\pi)^d} \delta(\Omega - \Delta \epsilon_{\bm{k}}).
\end{equation}
Here, $\Delta \epsilon_{\bm{k}} = \epsilon^c_{\bm{k}} - \epsilon^v_{\bm{k}}$ is the energy difference between the conduction band and the valence band. Thus, the JDOS has a maximum at the frequency where the largest number of states can be excited. 
In this paper, to account for the broadening of the DOS due to dissipation, we substitute the $\delta$ function by the Lorentz function as $\delta(x) \sim \Gamma/(\pi^2 x^2 + \Gamma^2)$, where $\Gamma$ is the dissipation used in the calculation of Fig.~\ref{oa_normal_fig}.
We show the numerical results of the JDOS in the left panel of Fig.~\ref{jdos_fig}. The JDOS has a maximum around $\Omega = 2~\mathrm{eV}$, explaining the optical conductivity's peak. However, the JDOS cannot explain the peak shift of the optical activity.

\subsection{peak shift in optical activity}
This peak shift is unique to the orbital optical activity in metals. It becomes clear when the optical activity is represented by the Bloch band basis. The full representation is derived in a recent work \cite{10.21468/SciPostPhys.14.5.118} (see Eq.~(26) in the paper). 
It shows that there is a Fermi surface effect (This effect originates from a term including the derivative of the Fermi distribution function. This term only becomes finite in metals) affecting the interband transition besides the Fermi sea effect (arising from terms which do not include the derivative of the Fermi distribution function. These terms describe transitions between unoccupied and occupied states) as the antisymmetric part of the optical activity $\sigma^{(A)}_{\mu\nu\lambda}(\Omega)$ includes \cite{10.21468/SciPostPhys.14.5.118}
\begin{align}
    &M_{\mu\nu\lambda}(\Omega) \equiv \nonumber \\
    &\int_{\mathrm{BZ}} \frac{d^dk}{(2\pi)^d} \sum_{n \neq m} \frac{- \tilde{\Omega} \partial_{k_{\lambda}} f_{n\bm{k}} \epsilon_{mn\bm{k}}  }{\epsilon_{mn\bm{k}}^2 - \tilde{\Omega}^2 } \mathrm{Im} 
    \Bigl(
    \mathcal{A}^{\mu}_{nm\bm{k}} \mathcal{A}^{\nu}_{mn\bm{k}}
    \Bigr).
\end{align}
Here, we define $\epsilon_{mn\bm{k}} = \epsilon_{m\bm{k}} - \epsilon_{n\bm{k}}$, $\tilde{\Omega} = \Omega + i \Gamma$, and $\mathcal{A}^{\mu}_{nm\bm{k}} = i\braket{u_{n\bm{k}} | \partial_{k_{\mu}} u_{m\bm{k}} }$ is the Berry connection.
This term is unique to the orbital contribution and does not exist in the spin contribution, and it has been missed in previous works. Ref.~\cite{10.21468/SciPostPhys.14.5.118} has derived it, recently.
It demonstrates that the effect of optical transitions in the optical activity cannot be simply described by transitions from occupied to unoccupied states, which is described by the Fermi sea term.
We note that in this two-band model with particle-hole symmetry, the Fermi sea term is suppressed.
This Fermi surface term is dominant and is locally enhanced at the optical gap $2\mu$ because only the states satisfying $\epsilon_{n\bm{k}} \sim \mu$ contribute and only the state satisfying $\epsilon_{m\bm{k}} \sim -\mu$ due to the particle-hole symmetry, which is consistent with the result in Fig.~\ref{mudepend_fig}.

In addition, we obtain a sum rule related to the shifting peak structure from $M_{\mu\nu\lambda}(\Omega)$. The sum rule reads
\begin{align}
    &\int_0^{\infty} d\Omega \mathrm{Im} M_{\mu\nu\lambda} (\Omega) \nonumber \\
    &=
    \frac{\pi}{2} \varepsilon_{\mu\nu\theta} \int_{\mathrm{BZ}} \frac{d^dk}{(2\pi)^d} \sum_n f_{n\bm{k}} \partial_{k_{\lambda}} m^{\mathrm{orb},\theta}_{n\bm{k}}.
\end{align}
Here, we define the orbital magnetic moment $\bm{m}^{\mathrm{orb}}_{n\bm{k}} = \frac{1}{2} \mathrm{Im} \bra{ \bm{\nabla} u_{n\bm{k}} } \times (\epsilon_{n\bm{k}} - H_{\bm{k}} ) \ket{\bm{\nabla} u_{n\bm{k}} }$.
This sum rule means that the area of this shifting peak structure is determined by the OMMD.

\subsection{Optical rotation} \label{sec_kerr_effect}

The optical activity $\sigma_{xyx}$ in our model can be observed using the optical rotation, the phenomenon in which the polarized axis of linearly polarized light rotates. 
The optical rotation is usually caused by the optical Hall effect in magnets without time-reversal symmetry, which is called the Kerr effect and the Faraday effect. Because the present model obeys time-reversal symmetry, the optical rotation originating from the optical Hall effect does not occur.
However, the optical activity also generates an optical rotation by a different mechanism, as described in the following discussion.

The spectrum of the longitudinal optical conductivity $\sigma_{ii}(\omega)$ can be extracted from the experimentally observed complex reflectance. Similarly, the spectrum of the optical activity can be extracted from the experimentally observed rotational angle and the ellipticity caused by the optical rotation. 
In this subsection, we discuss the optical rotation as a possibility of observing the optical activity experimentally.

We consider an electromagnetic wave incident in the $xz$ plane at an angle of incidence of $\ang{45}$ (see Appendix \ref{app_kerr} for a detailed discussion and calculation). The two-dimensional material is put on the $z=0$ plane, and the incident electric field is assumed to be $s$-polarized, i.e., pointing in the $y$-axis direction. Then, the reflected electric field, $(E^{\mathrm{r}}_y, E^{\mathrm{r}}_{xz})$, and the transmitted electric field, $(E^{\mathrm{t}}_y, E^{\mathrm{t}}_{xz})$, satisfy the following ratio
\begin{gather}
    \frac{E^{\mathrm{r}}_{xz}}{E^{\mathrm{r}}_y} \sim \frac{\Omega \sigma_{xyx}}{ c (\mu_0 c \sigma_{yy})} \biggl( \frac{2\sqrt{2}}{\mu_0 c} + \sigma_{xx}  \biggr)^{-1}, \label{kerr_ref} \\
    \frac{E^{\mathrm{t}}_{xz}}{E^{\mathrm{t}}_y} = -\frac{ \Omega \sigma_{xyx}}{\sqrt{2} c} \biggl( \frac{2\sqrt{2}}{\mu_0 c} + \sigma_{xx} \biggr)^{-1}. \label{kerr_tra}
\end{gather}
Here, we ignore the squared term of $\sigma_{xyx}$ in Eq.~(\ref{kerr_ref}) because it is very small. $\mu_0$ and $c$ are the permeability of the vacuum and the light velocity, and the product of them $(\mu_0 c)^{-1} \sim 10.90 e^2/ \hbar$ is the inverse of the vacuum impedance.
Usually, a two-dimensional material is attached to a substrate. We assume that the refractive index of the substrate at the frequency of interest is 1 in this paper. The component $E^{\mathrm{r},\mathrm{t}}_{xz}$ is the transverse response induced by the magnetoelectric effect $\chi_{xz}=-i\sigma_{xyx}$, and it is perpendicular to the usual response $E^{\mathrm{r},\mathrm{t}}_y$. Thus, the output electric field is no longer linearly polarized, but, in general, it becomes an ellipse whose polarised axis is tilted from the $y$ axis. 
The ratio relations in Eqs.~(\ref{kerr_ref}) and (\ref{kerr_tra}) give the ratio of the magnitude and the phase difference between electric fields perpendicular to each other. They determine the rotational angle $\theta$ and the ellipticity $\varepsilon$ of the output electric fields (see Appendix \ref{app_kerr} for the equation of $\theta$ and $\varepsilon$).
We show the rotational angle and the ellipticity in the middle and right panels in Fig.~\ref{jdos_fig} using the data of Fig.~\ref{oa_normal_fig}. The optical rotation is enhanced at $\Omega = 1~\mathrm{eV}$ corresponding to the peak of the optical activity. 
Its magnitude reaches about 2~\textmu$\mathrm{rad}$ for the reflected light and $20~\mathrm{nrad}$ for the transmitted one.
We see that in the low-frequency regime, the ellipticity predominates because the optical conductivity is dominated by the real part, and the optical activity is dominated by the imaginary part.
In other words, the phases of these two electric fields exhibit a deviation of nearly $\pi/2$  from each other.
Its magnitude decreases, reaching about $\varepsilon^{\mathrm{r}} \sim 1~\mathrm{nrad}$ and $\varepsilon^{\mathrm{t}} \sim 0.05~\mathrm{nrad}$ at $\Omega = 1~\mathrm{meV}$.

Next, we comment on the order of the ellipticity for real two-dimensional materials such as doped monolayer TMDs and TBG in the low-frequency regime. The conductivity of TMDs and TBG is $10 e^2/\hbar$ at the maximum \cite{radisavljevic2013mobility,ma2019experimental,polshyn2019large,lu2019superconductors}. Thus, it is smaller or comparable to $(\mu_0 c)^{-1}$. Therefore, the ellipticity of the reflection light in this regime is approximately written as $k \chi'/\sigma'$, where $k$ is the wavenumber of the light, $\chi'$ is the real part of $\chi_{\mu \nu}$, and $\sigma'$ is the real part of the conductivity. This equation is the Edelstein coefficient $\alpha =\chi'/\sigma'$ multiplied by the wavenumber. For strained $\mathrm{MoS_2}$ with $1\%$ strain, the Edelstein coefficient is observed to be $\alpha = 0.1~\mathrm{\AA}$ \cite{PhysRevLett.123.036806}. Thus, the ellipticity is converted to $\varepsilon^{\mathrm{r}} = 8~\mathrm{nrad}$ for the frequency $\Omega = 1~\mathrm{meV}$ (i.e. $k = 8~\mathrm{cm^{-1}}$), which is comparable to our calculation. For TBG \cite{PhysRevResearch.3.L032012}, the Edelstein coefficient is expected to be larger, reaching $10~\mathrm{\AA}$. Thus, the ellipticity will be 0.8~\textmu rad at $\Omega = 1~\mathrm{meV}$.

\section{Model calculation in the superconducting state} \label{model_sc}
In this section, we discuss the optical activity in the superconducting state. We use the same  Hamiltonian $H_{\bm{k}}$ mimicking strained $\mathrm{MoS_2}$ as in Sec.~\ref{model_normal}. We assume that the superconducting state is a uniform singlet. Thus, we introduce the pair potential as $\Delta_{\bm{k}} = \Delta \tau_0 \otimes i\sigma_y$, where $\Delta$ is real and $\tau_0$ is the identity matrix for the sublattice degrees of freedom. When calculating the optical conductivity and the optical activity using Green's functions (Eq.~\ref{oa_green_sc}), we introduce the dissipation effect phenomenologically. For this purpose, we introduce the renormalization factor 
\begin{equation}
\eta_{\omega} = 1 + \Gamma \biggr( \frac{\theta(|\Delta| - |\omega|)}{\sqrt{\Delta^2 - \omega^2}} + \frac{i \mathrm{sign}(\omega) \theta(|\omega|-|\Delta|)}{\sqrt{\omega^2 - \Delta^2}} \biggr),
\end{equation}
where $\theta(x)$ is the step function and $\mathrm{sign}(x)$ is the sign function returning $+1$ if $x>0$ and $-1$ if $x<0$. Then, we treat the dissipation effect by impurities as replacing $\omega$ and $\Delta$ in the Green's function $G^{R/A}_{\mathrm{BdG}}(\bm{k},\omega)$ by $\eta_{\omega} \omega$ and $\eta_{\omega} \Delta$. This renormalization is derived from the Born approximation \cite{kopnin2001nonequilibrium}. In the limit $\Delta \to 0$, it is consistent with a constant dissipation $i\Gamma$ used in the calculations in the normal state.

\begin{figure}[t]
\includegraphics[width=0.49\linewidth]{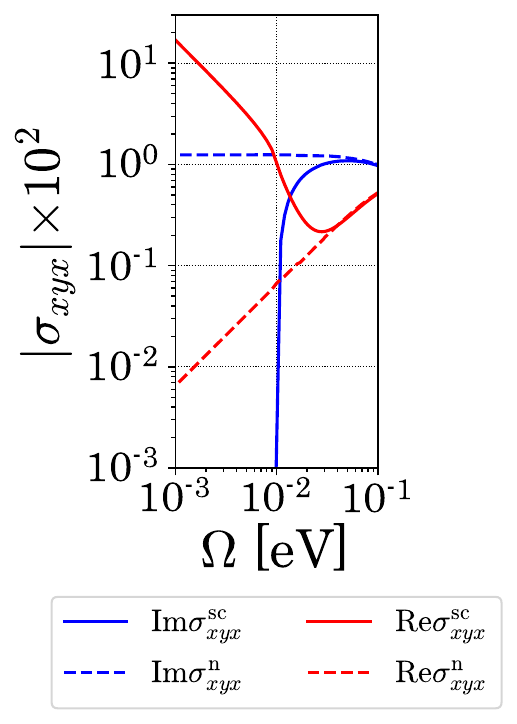}
\includegraphics[width=0.49\linewidth]{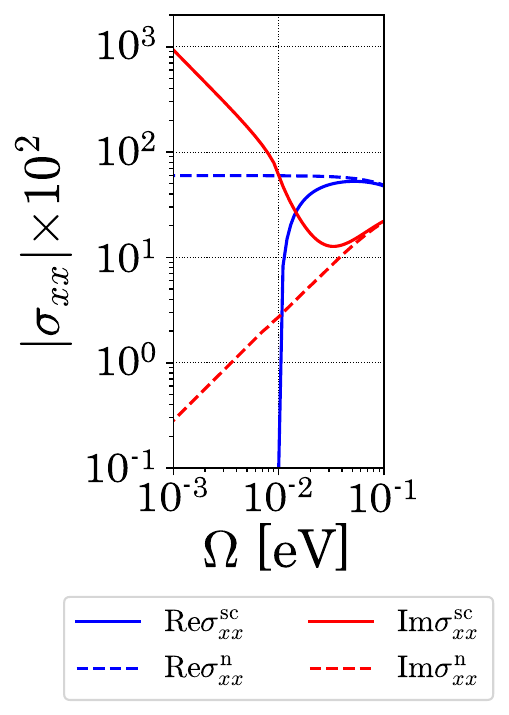}
\caption{Numerical results of the optical activity $\sigma_{xyx}$ (Left) and the optical conductivity $\sigma_{xx}$ (Right) in the superconducting state on a logarithmic scale. We denote the superconducting state and the normal state as sc and n, respectively. We use the same parameters and units as in Fig.~\ref{oa_normal_fig}. We set $\Delta = 0.005$.} \label{oa_sc_fig}
\end{figure}

We show the numerical results of the optical activity $\sigma_{xyx}$ and the optical conductivity $\sigma_{xx}$ of the superconducting state in Fig.~\ref{oa_sc_fig} (solid lines). We also include the results of the normal state for comparison (dashed lines). The spectra differ from the normal state below $\Omega = 2\Delta = 0.01$. $\mathrm{Im} \sigma_{xyx}$ and $\mathrm{Re} \sigma_{xx}$ decay rapidly with decreasing frequency around $2\Delta$, and the weights become zero in the gap. This difference in the spectra from the normal state is called the missing area. $\mathrm{Re} \sigma_{xyx}$ and $\mathrm{Im} \sigma_{xx}$ are enhanced in the gap and follow a $\Omega^{-1}$ dependence. The missing area and the anomalous divergence indicate the existence of a condensation. In particular, the missing area of the optical conductivity is known as the Ferrell-Glover-Tinkham sum rule for measuring the superfluid density \cite{PhysRev.109.1398,PhysRevLett.2.331}.
Similarly, they can be used to measure the SEE by the optical activity as discussed in Sec.~\ref{SEE}.

\begin{figure}[t]
\includegraphics[width=0.49\linewidth]{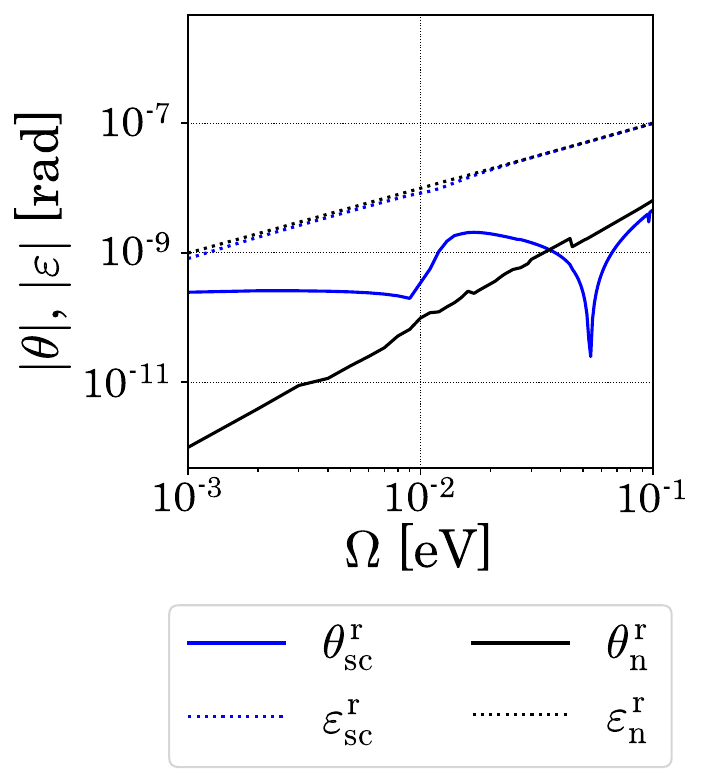}
\includegraphics[width=0.49\linewidth]{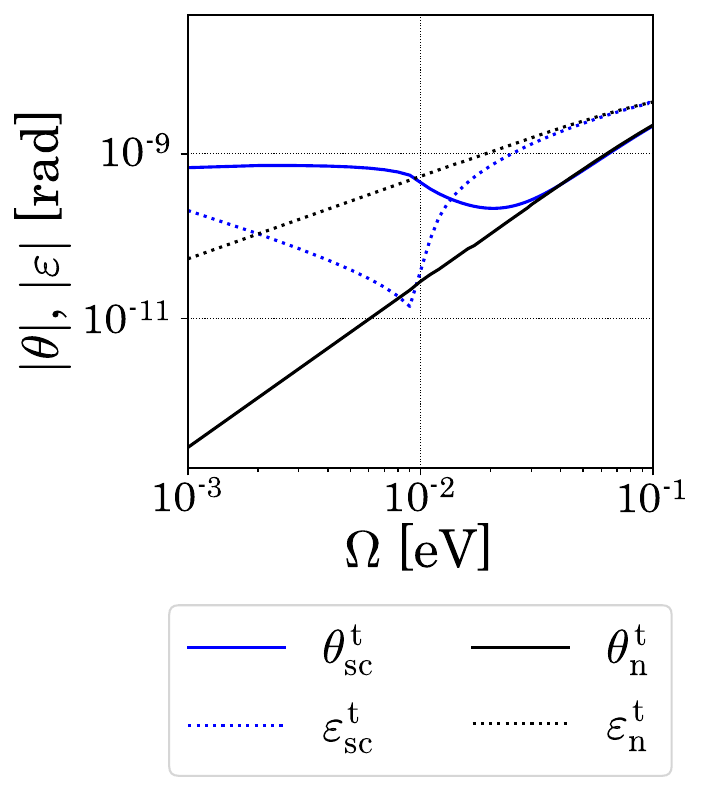}
\caption{Numerical results of the rotational angle $\theta$ and ellipticity $\varepsilon$ for the reflected light (Left) and the transmitted light (Right) on a logarithmic scale. We denote the superconducting state and the normal state as sc and n, respectively. We use the same parameters and units as in Fig.~\ref{oa_normal_fig}. We set $\Delta = 0.005$.} \label{kerr_sc_fig}
\end{figure}

The optical rotation is also changed because of the spectral modification by the superconducting state. We show the rotational angle $\theta$ and ellipticity $\varepsilon$ in Fig.~\ref{kerr_sc_fig} (blue lines). We also show the results of the normal state for comparison (black lines). As discussed in Sec.~\ref{sec_kerr_effect}, in the normal state, Eqs.~(\ref{kerr_ref}) and (\ref{kerr_tra}) are almost purely imaginary in the Drude regime. Thus, the phase difference of the electric fields is almost $\pi/2$, and the ellipticity is dominant. In the superconducting state, for frequencies inside the superconducting gap, the dominant components of $\sigma_{xx/yy}$ and $\sigma_{xyx}$ are changed to the imaginary part and the real part, respectively. Therefore, Eq.~(\ref{kerr_ref}) is no longer purely imaginary, and the rotational angle is enhanced. At low frequencies, $\mathrm{Im} \sigma_{xx}$ increases by $\Omega^{-1}$ and becomes comparable to the inverse vacuum impedance $(\mu_0 c)^{-1}$. Then, the real part and imaginary part of Eq.~(\ref{kerr_ref}) are comparable, and the rotational angle and ellipticity are in the same order of magnitude for the reflected light. For transmitted light, Eq.~(\ref{kerr_tra}) is also no longer purely imaginary and becomes rather real when $(\mu_0c)^{-1} > \mathrm{Im} \sigma_{xx}$, thus the rotational angle is dominant. As the frequency decreases further, $\mathrm{Im} \sigma_{xx}$ increases by $\Omega^{-1}$ and becomes comparable to $(\mu_0c)^{-1}$. Then, the rotational angle and the ellipticity also become comparable in magnitude.

\section{Conclusion and discussion} 
\subsection{Conclusion} \label{conclusion}
In this paper, we have discussed the optical activity originating from the orbital magnetic moment in metals and superconductors using a Green's function formalism. 
We have reconfirmed that the no-go theorem in the normal state, stating that a singular contribution to the optical activity is absent, is valid except for the chiral magnetic effect and, we have shown that this theorem is broken for the superconducting state in Secs.~\ref{oa_normal_sec} and \ref{oa_sc_sec}.
Then, this singularity contributes to the $\delta$ function in $\mathrm{Im} \sigma_{\mu\nu\lambda}$ in the DC-limit and the anomalous $\Omega^{-1}$-divergence in $\mathrm{Re} \sigma_{\mu \nu \lambda}$ for superconductors.
In the superconducting state, the coefficient of the singular part is confirmed to be exactly the superconducting Edelstein effect (SEE), which implies the validity of the missing area measurement for the SEE. 
An identical discussion was recently presented for the spin-induced optical activity. In this paper, it has been shown that the same discussion holds for the orbital-induced optical activity. 
In Sec.~\ref{model_normal}, we have shown calculations using a model for monolayer TMDs in the normal state. 
We have used an effective two-band model to elucidate the typical behavior of the optical activity. We have revealed that the low-frequency spectrum is described by a Drude-like model and that the interband transition shows an atypical chemical potential dependence originating from the interband Fermi surface term, unique to the orbital part in metals. 
We have obtained a sum rule related to the spectrum of the interband Fermi surface term. It shows that the spectrum area is determined by the orbital magnetic moment dipole.
In addition, we have evaluated the spectrum of the optical rotation. In Sec.~\ref{model_sc}, we have calculated the spectrum of the optical activity in the superconducting state, and we have demonstrated the existence of the missing area and the anomalous divergence of the optical activity at low frequencies. In addition, we have discussed the modification of the optical rotation by the superconducting state.

\subsection{Discussion of possible future experiments and directions}
Finally, we discuss the possibility of an experimental realization separated into the interband regime and the intraband regime. As discussed in Sec.~\ref{model_normal}, the interband peak is located at about a few $\mathrm{eV}$. The rotational angle and ellipticity for the reflected light reach a few \textmu rad, which makes them detectable with the state-of-the-art instruments, having a resolution of 10 nrad \cite{PhysRevLett.97.167002,kapitulnik2009polar}. 
On the other hand, the resolution in the terahertz regime, corresponding to the gap scale of most noncentrosymmetric superconductors, \cite{tagay2023high} has recently been reported to reach 20 \textmu rad and is expected to go down to 2 \textmu rad.
The optical rotation of TBG is close to this technical observable limit. Thus, unless we can find other two-dimensional materials with a larger Edelstein coefficient $\alpha$, the experimental observation is currently challenging in this regime.

Finally, we shortly discuss the three-dimensional materials. In three-dimensional systems, the optical rotation caused by light transmitted through materials is expected. 
As light cannot easily penetrate superconductors, transmission can only occur in thin films \cite{PhysRevLett.111.057002}.
This effect originates from the difference of the refractive index for left and right circular polarized lights $\Delta n$. It is given by $\Phi ( = \theta + i\varepsilon ) = - \Delta n k \zeta /2$, where $\theta$ and $\varepsilon$ are the rotational angle and the ellipticity, $k$ is the wavenumber of the light, and $\zeta$ is the thickness of materials. For the natural optical activity, the refractive index difference is given by $\Delta n = \mu_0 c \chi$ \cite{raab2004multipole}; thus, we can directly obtain the spectrum of the optical activity $\chi$ from the optical rotation spectrum. In the case of noncentrosymmetric cubics (point group: O and T), the optical activity in the THz regime is $\chi \sim (e \mu_{\mathrm{B}} \tau k_F^2/\hbar) (\gamma k_F/E_F)$ with an antisymmetric SOC term $\gamma \bm{k} \cdot \bm{\sigma}$. Here, $\mu_{\mathrm{B}}$ is the Bohr magneton, $\tau$ is the relaxation time, and $k_F$ and $E_F$ are the Fermi wavenumber and the Fermi energy. Candidates for superconductors with this SOC include $\mathrm{Li_2Pd_3B}$, $\mathrm{Li_2Pt_3B}$, and $\mathrm{Mo_2Al_3C}$ \cite{smidman2017superconductivity}.
Then, the optical rotation reads
\begin{equation}
\Phi \sim - \frac{1}{2}\frac{e\mu_{\mathrm{B}} \mu_0 k_F }{ \hbar} (\Omega \tau)(k_F \zeta) \frac{\gamma k_F }{E_F},
\end{equation}
and the typical order of magnitude is estimated to be 4.8~\textmu rad for $k_F = 1~\mathrm{\AA}^{-1}$, $\tau = 10~\mathrm{fs}$, and $\gamma k_F/E_F = 10^{-2}$ at $\Omega = 1~\mathrm{THz}$ using the thickness $\zeta = 500~\mathrm{\AA}$.
It indicates that this optical rotation is observable in the THz regime \cite{tagay2023high}.

\section*{acknowledgements}
K.S. acknowledges support as a JSPS research fellow and is supported by JSPS KAKENHI, Grant No.22J23393 and No.22KJ2008. 
R.P. is supported by JSPS KAKENHI No.23K03300.

\appendix
\begin{widetext}
\section{No-go theorem} \label{nogo}
The singular part of the optical activity is restricted in the normal state. This part was shown to be forbidden for the spin contribution \cite{PhysRevB.108.165119}. In this appendix, we will show that the part for the orbital part is also restricted. The coefficient of the singular part is
\begin{align}
    \Phi_{\mu \nu \lambda}(0)
    &=
    \frac{e^2}{2} \int [d^4k] f(\omega) \mathrm{Tr}
    \Bigl[
    -\partial_{\lambda}G^{RA}(\bm{k},\omega) v_{\bm{k}}^{\mu} G^{R}(\bm{k},\omega) v_{\bm{k}}^{\nu}
    +G^{RA}(\bm{k},\omega) v_{\bm{k}}^{\mu} \partial_{\lambda} G^R(\bm{k},\omega) v_{\bm{k}}^{\nu}
    \nonumber \\
    & \hspace{90pt} -\partial_{\lambda} G^A(\bm{k},\omega) v_{\bm{k}}^{\mu} G^{RA}(\bm{k},\omega) v^{\nu}_{\bm{k}}
    +G^A(\bm{k},\omega) v_{\bm{k}}^{\mu} \partial_{\lambda} G^{RA}(\bm{k},\omega) v_{\bm{k}}^{\nu}  
    \Bigr] \nonumber \\
    &=
    \frac{e^2}{2} \int [d^4k] f(\omega) \mathrm{Tr}
    \Bigl[
    -\partial_{\lambda}G^{R}(\bm{k},\omega) v_{\bm{k}}^{\mu} G^{R}(\bm{k},\omega) v_{\bm{k}}^{\nu}
    +G^{R}(\bm{k},\omega) v_{\bm{k}}^{\mu} \partial_{\lambda} G^R(\bm{k},\omega) v_{\bm{k}}^{\nu}
    \nonumber \\
    & \hspace{90pt} +\partial_{\lambda} G^A(\bm{k},\omega) v_{\bm{k}}^{\mu} G^{A}(\bm{k},\omega) v^{\nu}_{\bm{k}}
    -G^A(\bm{k},\omega) v_{\bm{k}}^{\mu} \partial_{\lambda} G^{A}(\bm{k},\omega) v_{\bm{k}}^{\nu}  
    \Bigr] \nonumber \\
    &=
    e^2 \int [d^4k] f(\omega) \mathrm{Re} \mathrm{Tr}
    \Bigl[
    -\partial_{\lambda}G^{R}(\bm{k},\omega) v_{\bm{k}}^{\mu} G^{R}(\bm{k},\omega) v_{\bm{k}}^{\nu}
    +G^{R}(\bm{k},\omega) v_{\bm{k}}^{\mu} \partial_{\lambda} G^R(\bm{k},\omega) v_{\bm{k}}^{\nu} \Bigr].
\end{align}
Here, $\int [d^4k] = \int^{\infty}_{-\infty} d\omega /(2\pi i) \int_{\mathrm{BZ}} d^3 k /(2\pi)^3$. Thus, this equation is purely imaginary. Using the relation $\partial_{\lambda} G^{R}(\bm{k},\omega) = G^{R}(\bm{k},\omega) v^{\lambda}_{\bm{k}} G^{R}(\bm{k},\omega)$, we obtain
\begin{align}
    \Phi_{\mu \nu \lambda}(0)
    &=
    e^2 \int [d^4k] f(\omega) \mathrm{Re} \mathrm{Tr}
    \Bigl[
    -G^{R}(\bm{k},\omega) v^{\lambda}_{\bm{k}} G^{R}(\bm{k},\omega) v_{\bm{k}}^{\mu} G^{R}(\bm{k},\omega) v_{\bm{k}}^{\nu}
    +G^{R}(\bm{k},\omega) v_{\bm{k}}^{\mu} G^R(\bm{k},\omega) v^{\lambda}_{\bm{k}} G^{R}(\bm{k},\omega) v_{\bm{k}}^{\nu} \Bigr] \nonumber \\
    &=
    e^2 \varepsilon_{\mu \nu \lambda} \int [d^4k] f(\omega) \mathrm{Re} \mathrm{Tr}
    \Bigl[
    -G^{R}(\bm{k},\omega) v^{x}_{\bm{k}} G^{R}(\bm{k},\omega) v_{\bm{k}}^{y} G^{R}(\bm{k},\omega) v_{\bm{k}}^{z}
    +G^{R}(\bm{k},\omega) v_{\bm{k}}^{y} G^R(\bm{k},\omega) v^{x}_{\bm{k}} G^{R}(\bm{k},\omega) v_{\bm{k}}^{z} \Bigr].
\end{align}
This equation is totally antisymmetric. Then, we show that most components vanish. 
Only the totally antisymmetric part remains. This remaining part is known to be the chiral magnetic effect, as mentioned in the main text. It can be shown that this part is also zero by its topological nature using the band representation \cite{PhysRevLett.116.077201,10.21468/SciPostPhys.14.5.118}.

In the superconducting case, this no-go theorem changes. Calculating the singular term in the same way, we obtain
\begin{align}
    \Phi_{\mu \nu \lambda}(0) &=
    \frac{e^2}{2} \int [d^4k] f(\omega) \mathrm{Re} \mathrm{Tr} \Bigl[
    -\partial_{\lambda} G^{R}_{\mathrm{BdG}}(\bm{k},\omega) \tilde{v}^{\mu}_{\bm{k}} G^{R}_{\mathrm{BdG}}(\bm{k},\omega) \tilde{v}^{\nu}_{\bm{k}}
    +G^{R}_{\mathrm{BdG}}(\bm{k},\omega) \tilde{v}^{\mu}_{\bm{k}} \partial_{\lambda} G^R_{\mathrm{BdG}}(\bm{k},\omega) \tilde{v}^{\nu}_{\bm{k}} \Bigr].
\end{align}
This equation is purely imaginary and antisymmetric under the interchange between $\mu$ and $\nu$. There is one difference to the case of the normal state. The velocity operator appearing by differentiating the Green's function $\hat{v}^{\mu}_{\bm{k}}$ is different from the  velocity operator $\tilde{v}^{\mu}_{\bm{k}}$. Thus, many components of the singular part can be finite.
\end{widetext}

\section{Optical rotation} \label{app_kerr}
Here, we derive the rotational angle in two-dimensional metals with an in-plane polar axis pointing in the $x$ direction. In this case, the in-plane current induces the out-of-plane magnetization, which is often discussed in strained TMDs and twisted bilayer graphene. In the following, we consider that the two-dimensional metal is put on an insulating substrate, which has a real-valued refractive index $n_{\mathrm{sub}}$ in the frequency regime of interest, and we assume $n_{\mathrm{sub}} \sim 1$.
\begin{figure}[t]
\includegraphics[width=0.8\linewidth]{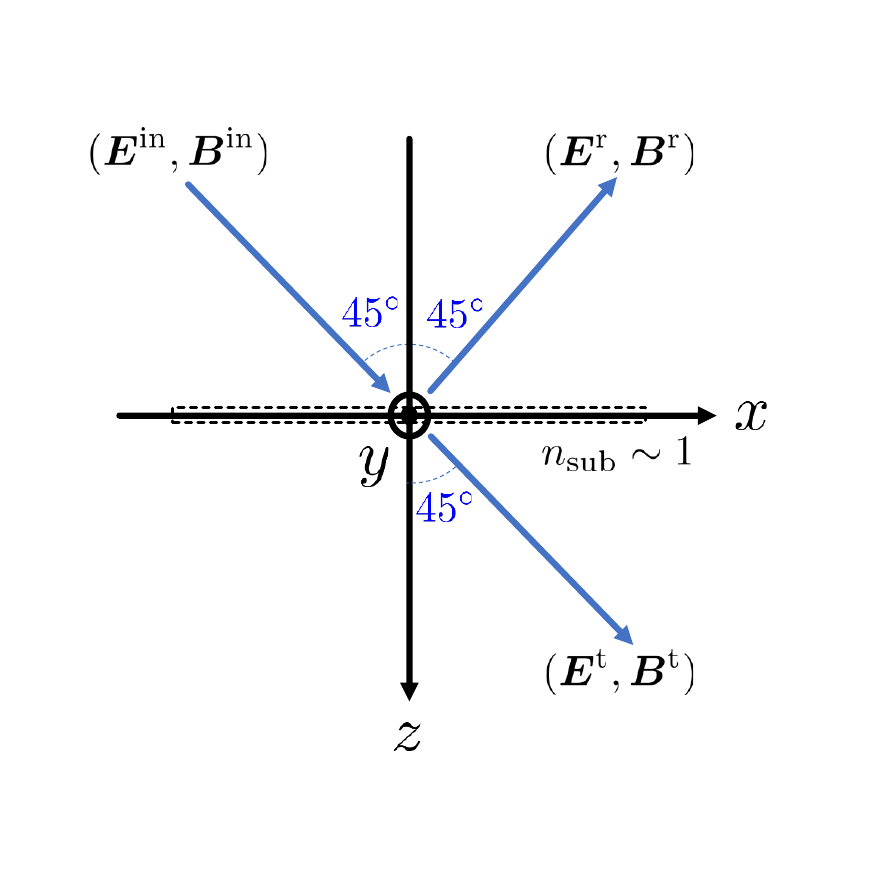}
\caption{Schematic diagram visualizing the incident, reflected, and transmitted light beam. The two-dimensional material is put on the $xy$ plane (dashed line).}  \label{snell_fig}
\end{figure}

We will derive the rotational angle using Maxwell's electrodynamics. We consider that the incident light is in the $xz$ plane and travels at an angle of incidence of $\ang{45}$. Furthermore, the electric field of the incident light is the $s$-polarized and, thus, is oriented in the $y$ direction. In this case, the electric field for the incident light $\bm{E}^{\mathrm{in}}$, the reflected light $\bm{E}^{\mathrm{r}}$, and the refracted light $\bm{E}^{\mathrm{t}}$ are given by
\begin{subequations}
  \begin{align}
    &\bm{E}^{\mathrm{in}} = E^{\mathrm{in}} \bm{e}_y e^{i \bm{k} \cdot \bm{r} - i \omega t}, \\
    &\bm{E}^{\mathrm{r}} = (E^{\mathrm{r}}_y  \bm{e}_y + E^{\mathrm{r}}_{xz} (\bm{e}_x + \bm{e}_z) ) e^{i \bm{k}' \cdot \bm{r} - i \omega t}, \\
    &\bm{E}^{\mathrm{t}} = (E^{\mathrm{t}}_y  \bm{e}_y + E^{\mathrm{t}}_{xz} (\bm{e}_x - \bm{e}_z) ) e^{i \bm{k} \cdot \bm{r} - i \omega t}.
  \end{align}
\end{subequations}
Here, the wavenumbers are given by $\bm{k} = \frac{k}{\sqrt{2}} ( \bm{e}_x + \bm{e}_z)$ and $\bm{k}' = \frac{k}{\sqrt{2}} ( \bm{e}_x - \bm{e}_z)$ and the magnitude of the wavenumber $k$ satisfies the dispersion relation $\omega  = c k$, where $\omega$ is the frequency and $c$ is the light velocity. The corresponding magnetic fields are determined by Faraday's law ($\bm{\nabla} \times \bm{E} = -\partial \bm{B} / \partial t$). They read
\begin{subequations}
  \begin{align}
    &\bm{B}^{\mathrm{in}} = \frac{E^{\mathrm{in}}}{\sqrt{2}c} (\bm{e}_z - \bm{e}_x) e^{i \bm{k} \cdot \bm{r} - i \omega t}, \\
    &\bm{B}^{\mathrm{r}} = \frac{1}{\sqrt{2 }c}( -2E^{\mathrm{r}}_{xz} \bm{e}_y + E^{\mathrm{r}}_y (\bm{e}_x +\bm{e}_z) ) e^{i \bm{k}' \cdot \bm{r} - i \omega t}, \\
    &\bm{B}^{\mathrm{t}} = \frac{1}{\sqrt{2 }c}( 2E^{\mathrm{t}}_{xz} \bm{e}_y + E^{\mathrm{t}}_y (\bm{e}_z -\bm{e}_x) ) e^{i \bm{k} \cdot \bm{r} - i \omega t}.
  \end{align}
\end{subequations}

Electrons are driven by the electromagnetic wave in the two-dimensional material, and we treat the responses up to the linear order. In the following, we consider that the linear responses only include a usual longitudinal conductivity $\sigma$ and a magnetoelectric effect $\chi$. These effects appear in a current density as
\begin{align}
    j_x &= \sigma_{xx} (\bm{E}^{\mathrm{t}})_x + ik_x \chi (\bm{E}^{\mathrm{t}})_y \nonumber \\ 
    &=
    \sigma_{xx} E^{\mathrm{t}}_{xz} + ik_x \chi E^{\mathrm{t}}_y, \\
    j_y &= \sigma_{yy} (\bm{E}^{\mathrm{t}})_y - ik_x \chi (\bm{E}^{\mathrm{t}})_x \nonumber \\
    &=
    \sigma_{yy} E^{\mathrm{t}}_y - ik_x \chi E^{\mathrm{t}}_{xz}.
\end{align}
This current flows in these two-dimensional metals and influences the boundary conditions of the electromagnetic fields. The boundary conditions coming from Faraday's law and Gauss's law are given by
\begin{equation}
    E^{\mathrm{in}} + E^{\mathrm{r}}_y = E^{\mathrm{t}}_y, \hspace{10pt}E^{\mathrm{r}}_{xz} = E^{\mathrm{t}}_{xz}.
\end{equation}
Furthermore, Amp\`ere's law gives two boundary conditions as
\begin{subequations}
  \begin{gather}
    \frac{-1}{\sqrt{2} \mu_0 c} (2E^{\mathrm{r}}_{xz} + 2E^{\mathrm{t}}_{xz} ) = \sigma_{xx} E^{\mathrm{t}}_{xz} + i \frac{\omega}{\sqrt{2}c} \chi E^{\mathrm{t}}_{y}, \\
    \frac{1}{\sqrt{2} \mu_0 c} (E^{\mathrm{in}} - E^{\mathrm{r}}_{y} - E^{\mathrm{t}}_y) = \sigma_{yy} E^{\mathrm{t}}_y - i \frac{\omega}{\sqrt{2}c} \chi E^{\mathrm{t}}_{xz}.
  \end{gather}
\end{subequations}
Here, $\mu_0$ is the vacuum permeability.
Solving these equations simultaneously, we find
\begin{align}
    \frac{E^{\mathrm{t}}_{xz}}{E^{\mathrm{t}}_y} &= -\frac{i \omega \chi}{\sqrt{2} c} \biggl( \frac{2\sqrt{2}}{\mu_0 c} + \sigma_{xx} \biggr)^{-1} \equiv t e^{i\phi_t}, \\
    \frac{E^{\mathrm{r}}_{xz}}{E^{\mathrm{r}}_y} &= \frac{i \omega \chi}{ c (\mu_0 c \sigma_{yy})} \biggl( \frac{2\sqrt{2}}{\mu_0 c} + \sigma_{xx} - \frac{\omega^2 \chi^2}{2c^2 \sigma_{yy}} \biggr)^{-1} \nonumber \\
    &\sim \frac{i \omega \chi}{ c (\mu_0 c \sigma_{yy})} \biggl( \frac{2\sqrt{2}}{\mu_0 c} + \sigma_{xx}  \biggr)^{-1} \equiv r e^{i\phi_r}.
\end{align}
Here, we ignore the square of $\chi$ due to its smallness.
This ratio gives the rotational angle $\theta$ and ellipticity $\varepsilon$ as
\begin{subequations}
    \begin{align}
        \theta^r &= \frac{1}{2} \arctan \biggl( \frac{2r \cos \phi_r }{1-r^2}  \biggr), \\
        \varepsilon^r &= \tan \bigg\{ \frac{1}{2} \arcsin \biggl( \frac{2r \sin \phi_r }{1+r^2}  \biggr)\bigg\}.
    \end{align}
\end{subequations}
The rotational angle and ellipticity of the transmitted light are obtained by simply replacing $r$ with $t$.

\bibliography{reference.bib}

\begin{thebibliography}{75}%
\makeatletter
\providecommand \@ifxundefined [1]{%
 \@ifx{#1\undefined}
}%
\providecommand \@ifnum [1]{%
 \ifnum #1\expandafter \@firstoftwo
 \else \expandafter \@secondoftwo
 \fi
}%
\providecommand \@ifx [1]{%
 \ifx #1\expandafter \@firstoftwo
 \else \expandafter \@secondoftwo
 \fi
}%
\providecommand \natexlab [1]{#1}%
\providecommand \enquote  [1]{``#1''}%
\providecommand \bibnamefont  [1]{#1}%
\providecommand \bibfnamefont [1]{#1}%
\providecommand \citenamefont [1]{#1}%
\providecommand \href@noop [0]{\@secondoftwo}%
\providecommand \href [0]{\begingroup \@sanitize@url \@href}%
\providecommand \@href[1]{\@@startlink{#1}\@@href}%
\providecommand \@@href[1]{\endgroup#1\@@endlink}%
\providecommand \@sanitize@url [0]{\catcode `\\12\catcode `\$12\catcode
  `\&12\catcode `\#12\catcode `\^12\catcode `\_12\catcode `\%12\relax}%
\providecommand \@@startlink[1]{}%
\providecommand \@@endlink[0]{}%
\providecommand \url  [0]{\begingroup\@sanitize@url \@url }%
\providecommand \@url [1]{\endgroup\@href {#1}{\urlprefix }}%
\providecommand \urlprefix  [0]{URL }%
\providecommand \Eprint [0]{\href }%
\providecommand \doibase [0]{https://doi.org/}%
\providecommand \selectlanguage [0]{\@gobble}%
\providecommand \bibinfo  [0]{\@secondoftwo}%
\providecommand \bibfield  [0]{\@secondoftwo}%
\providecommand \translation [1]{[#1]}%
\providecommand \BibitemOpen [0]{}%
\providecommand \bibitemStop [0]{}%
\providecommand \bibitemNoStop [0]{.\EOS\space}%
\providecommand \EOS [0]{\spacefactor3000\relax}%
\providecommand \BibitemShut  [1]{\csname bibitem#1\endcsname}%
\let\auto@bib@innerbib\@empty
\bibitem [{\citenamefont {Barron}(2004)}]{barron_2004}%
  \BibitemOpen
  \bibfield  {author} {\bibinfo {author} {\bibfnamefont {L.~D.}\ \bibnamefont
  {Barron}},\ }\href {https://doi.org/10.1017/CBO9780511535468} {\emph
  {\bibinfo {title} {Molecular Light Scattering and Optical Activity}}},\
  \bibinfo {edition} {2nd}\ ed.\ (\bibinfo  {publisher} {Cambridge University
  Press},\ \bibinfo {year} {2004})\BibitemShut {NoStop}%
\bibitem [{\citenamefont {Landau}\ \emph {et~al.}(2013)\citenamefont {Landau},
  \citenamefont {Bell}, \citenamefont {Kearsley}, \citenamefont {Pitaevskii},
  \citenamefont {Lifshitz},\ and\ \citenamefont
  {Sykes}}]{landau2013electrodynamics}%
  \BibitemOpen
  \bibfield  {author} {\bibinfo {author} {\bibfnamefont {L.~D.}\ \bibnamefont
  {Landau}}, \bibinfo {author} {\bibfnamefont {J.~S.}\ \bibnamefont {Bell}},
  \bibinfo {author} {\bibfnamefont {M.}~\bibnamefont {Kearsley}}, \bibinfo
  {author} {\bibfnamefont {L.}~\bibnamefont {Pitaevskii}}, \bibinfo {author}
  {\bibfnamefont {E.}~\bibnamefont {Lifshitz}},\ and\ \bibinfo {author}
  {\bibfnamefont {J.}~\bibnamefont {Sykes}},\ }\href@noop {} {\emph {\bibinfo
  {title} {Electrodynamics of Continuous Media}}},\ Vol.~\bibinfo {volume} {8}\
  (\bibinfo  {publisher} {Elsevier},\ \bibinfo {year} {2013})\BibitemShut
  {NoStop}%
\bibitem [{\citenamefont {Pisarev}\ \emph {et~al.}(1991)\citenamefont
  {Pisarev}, \citenamefont {Krichevtsov},\ and\ \citenamefont
  {Pavlov}}]{pisarev1991optical}%
  \BibitemOpen
  \bibfield  {author} {\bibinfo {author} {\bibfnamefont {R.}~\bibnamefont
  {Pisarev}}, \bibinfo {author} {\bibfnamefont {B.}~\bibnamefont
  {Krichevtsov}},\ and\ \bibinfo {author} {\bibfnamefont {V.}~\bibnamefont
  {Pavlov}},\ }\bibfield  {title} {\bibinfo {title} {Optical study of the
  antiferromagnetic-paramagnetic phase transition in chromium oxide cr2o3},\
  }\href {https://www.tandfonline.com/doi/abs/10.1080/01411599108203448}
  {\bibfield  {journal} {\bibinfo  {journal} {Phase Transitions: A
  Multinational Journal}\ }\textbf {\bibinfo {volume} {37}},\ \bibinfo {pages}
  {63} (\bibinfo {year} {1991})}\BibitemShut {NoStop}%
\bibitem [{\citenamefont {Krichevtsov}\ \emph {et~al.}(1993)\citenamefont
  {Krichevtsov}, \citenamefont {Pavlov}, \citenamefont {Pisarev},\ and\
  \citenamefont {Gridnev}}]{krichevtsov1993spontaneous}%
  \BibitemOpen
  \bibfield  {author} {\bibinfo {author} {\bibfnamefont {B.}~\bibnamefont
  {Krichevtsov}}, \bibinfo {author} {\bibfnamefont {V.}~\bibnamefont {Pavlov}},
  \bibinfo {author} {\bibfnamefont {R.}~\bibnamefont {Pisarev}},\ and\ \bibinfo
  {author} {\bibfnamefont {V.}~\bibnamefont {Gridnev}},\ }\bibfield  {title}
  {\bibinfo {title} {Spontaneous non-reciprocal reflection of light from
  antiferromagnetic cr2o3},\ }\href
  {https://iopscience.iop.org/article/10.1088/0953-8984/5/44/014} {\bibfield
  {journal} {\bibinfo  {journal} {Journal of Physics: Condensed Matter}\
  }\textbf {\bibinfo {volume} {5}},\ \bibinfo {pages} {8233} (\bibinfo {year}
  {1993})}\BibitemShut {NoStop}%
\bibitem [{\citenamefont {Rikken}\ and\ \citenamefont
  {Raupach}(1997)}]{rikken1997observation}%
  \BibitemOpen
  \bibfield  {author} {\bibinfo {author} {\bibfnamefont {G.}~\bibnamefont
  {Rikken}}\ and\ \bibinfo {author} {\bibfnamefont {E.}~\bibnamefont
  {Raupach}},\ }\bibfield  {title} {\bibinfo {title} {Observation of
  magneto-chiral dichroism},\ }\href {https://www.nature.com/articles/37323}
  {\bibfield  {journal} {\bibinfo  {journal} {Nature}\ }\textbf {\bibinfo
  {volume} {390}},\ \bibinfo {pages} {493} (\bibinfo {year}
  {1997})}\BibitemShut {NoStop}%
\bibitem [{\citenamefont {Arima}(2008)}]{Arima_2008}%
  \BibitemOpen
  \bibfield  {author} {\bibinfo {author} {\bibfnamefont {T.}~\bibnamefont
  {Arima}},\ }\bibfield  {title} {\bibinfo {title} {Magneto-electric optics in
  non-centrosymmetric ferromagnets},\ }\href
  {https://doi.org/10.1088/0953-8984/20/43/434211} {\bibfield  {journal}
  {\bibinfo  {journal} {Journal of Physics: Condensed Matter}\ }\textbf
  {\bibinfo {volume} {20}},\ \bibinfo {pages} {434211} (\bibinfo {year}
  {2008})}\BibitemShut {NoStop}%
\bibitem [{\citenamefont {Fiebig}(2005)}]{fiebig2005revival}%
  \BibitemOpen
  \bibfield  {author} {\bibinfo {author} {\bibfnamefont {M.}~\bibnamefont
  {Fiebig}},\ }\bibfield  {title} {\bibinfo {title} {Revival of the
  magnetoelectric effect},\ }\href
  {https://iopscience.iop.org/article/10.1088/0022-3727/38/8/R01} {\bibfield
  {journal} {\bibinfo  {journal} {Journal of physics D: applied physics}\
  }\textbf {\bibinfo {volume} {38}},\ \bibinfo {pages} {R123} (\bibinfo {year}
  {2005})}\BibitemShut {NoStop}%
\bibitem [{\citenamefont {Pimenov}\ \emph {et~al.}(2006)\citenamefont
  {Pimenov}, \citenamefont {Mukhin}, \citenamefont {Ivanov}, \citenamefont
  {Travkin}, \citenamefont {Balbashov},\ and\ \citenamefont
  {Loidl}}]{pimenov2006possible}%
  \BibitemOpen
  \bibfield  {author} {\bibinfo {author} {\bibfnamefont {A.}~\bibnamefont
  {Pimenov}}, \bibinfo {author} {\bibfnamefont {A.}~\bibnamefont {Mukhin}},
  \bibinfo {author} {\bibfnamefont {V.~Y.}\ \bibnamefont {Ivanov}}, \bibinfo
  {author} {\bibfnamefont {V.}~\bibnamefont {Travkin}}, \bibinfo {author}
  {\bibfnamefont {A.}~\bibnamefont {Balbashov}},\ and\ \bibinfo {author}
  {\bibfnamefont {A.}~\bibnamefont {Loidl}},\ }\bibfield  {title} {\bibinfo
  {title} {Possible evidence for electromagnons in multiferroic manganites},\
  }\href {https://www.nature.com/articles/nphys212} {\bibfield  {journal}
  {\bibinfo  {journal} {Nature physics}\ }\textbf {\bibinfo {volume} {2}},\
  \bibinfo {pages} {97} (\bibinfo {year} {2006})}\BibitemShut {NoStop}%
\bibitem [{\citenamefont {Sushkov}\ \emph {et~al.}(2007)\citenamefont
  {Sushkov}, \citenamefont {Aguilar}, \citenamefont {Park}, \citenamefont
  {Cheong},\ and\ \citenamefont {Drew}}]{PhysRevLett.98.027202}%
  \BibitemOpen
  \bibfield  {author} {\bibinfo {author} {\bibfnamefont {A.~B.}\ \bibnamefont
  {Sushkov}}, \bibinfo {author} {\bibfnamefont {R.~V.}\ \bibnamefont
  {Aguilar}}, \bibinfo {author} {\bibfnamefont {S.}~\bibnamefont {Park}},
  \bibinfo {author} {\bibfnamefont {S.-W.}\ \bibnamefont {Cheong}},\ and\
  \bibinfo {author} {\bibfnamefont {H.~D.}\ \bibnamefont {Drew}},\ }\bibfield
  {title} {\bibinfo {title} {Electromagnons in multiferroic
  ${\mathrm{ymn}}_{2}{\mathrm{o}}_{5}$ and
  ${\mathrm{tbmn}}_{2}{\mathrm{o}}_{5}$},\ }\href
  {https://doi.org/10.1103/PhysRevLett.98.027202} {\bibfield  {journal}
  {\bibinfo  {journal} {Phys. Rev. Lett.}\ }\textbf {\bibinfo {volume} {98}},\
  \bibinfo {pages} {027202} (\bibinfo {year} {2007})}\BibitemShut {NoStop}%
\bibitem [{\citenamefont {Katsura}\ \emph {et~al.}(2007)\citenamefont
  {Katsura}, \citenamefont {Balatsky},\ and\ \citenamefont
  {Nagaosa}}]{PhysRevLett.98.027203}%
  \BibitemOpen
  \bibfield  {author} {\bibinfo {author} {\bibfnamefont {H.}~\bibnamefont
  {Katsura}}, \bibinfo {author} {\bibfnamefont {A.~V.}\ \bibnamefont
  {Balatsky}},\ and\ \bibinfo {author} {\bibfnamefont {N.}~\bibnamefont
  {Nagaosa}},\ }\bibfield  {title} {\bibinfo {title} {Dynamical magnetoelectric
  coupling in helical magnets},\ }\href
  {https://doi.org/10.1103/PhysRevLett.98.027203} {\bibfield  {journal}
  {\bibinfo  {journal} {Phys. Rev. Lett.}\ }\textbf {\bibinfo {volume} {98}},\
  \bibinfo {pages} {027203} (\bibinfo {year} {2007})}\BibitemShut {NoStop}%
\bibitem [{\citenamefont {Kida}\ \emph {et~al.}(2009)\citenamefont {Kida},
  \citenamefont {Okuyama}, \citenamefont {Ishiwata}, \citenamefont {Taguchi},
  \citenamefont {Shimano}, \citenamefont {Iwasa}, \citenamefont {Arima},\ and\
  \citenamefont {Tokura}}]{PhysRevB.80.220406}%
  \BibitemOpen
  \bibfield  {author} {\bibinfo {author} {\bibfnamefont {N.}~\bibnamefont
  {Kida}}, \bibinfo {author} {\bibfnamefont {D.}~\bibnamefont {Okuyama}},
  \bibinfo {author} {\bibfnamefont {S.}~\bibnamefont {Ishiwata}}, \bibinfo
  {author} {\bibfnamefont {Y.}~\bibnamefont {Taguchi}}, \bibinfo {author}
  {\bibfnamefont {R.}~\bibnamefont {Shimano}}, \bibinfo {author} {\bibfnamefont
  {K.}~\bibnamefont {Iwasa}}, \bibinfo {author} {\bibfnamefont
  {T.}~\bibnamefont {Arima}},\ and\ \bibinfo {author} {\bibfnamefont
  {Y.}~\bibnamefont {Tokura}},\ }\bibfield  {title} {\bibinfo {title}
  {Electric-dipole-active magnetic resonance in the conical-spin magnet
  ${\text{ba}}_{2}{\text{mg}}_{2}{\text{fe}}_{12}{\text{o}}_{22}$},\ }\href
  {https://doi.org/10.1103/PhysRevB.80.220406} {\bibfield  {journal} {\bibinfo
  {journal} {Phys. Rev. B}\ }\textbf {\bibinfo {volume} {80}},\ \bibinfo
  {pages} {220406} (\bibinfo {year} {2009})}\BibitemShut {NoStop}%
\bibitem [{\citenamefont {K\'ezsm\'arki}\ \emph {et~al.}(2011)\citenamefont
  {K\'ezsm\'arki}, \citenamefont {Kida}, \citenamefont {Murakawa},
  \citenamefont {Bord\'acs}, \citenamefont {Onose},\ and\ \citenamefont
  {Tokura}}]{PhysRevLett.106.057403}%
  \BibitemOpen
  \bibfield  {author} {\bibinfo {author} {\bibfnamefont {I.}~\bibnamefont
  {K\'ezsm\'arki}}, \bibinfo {author} {\bibfnamefont {N.}~\bibnamefont {Kida}},
  \bibinfo {author} {\bibfnamefont {H.}~\bibnamefont {Murakawa}}, \bibinfo
  {author} {\bibfnamefont {S.}~\bibnamefont {Bord\'acs}}, \bibinfo {author}
  {\bibfnamefont {Y.}~\bibnamefont {Onose}},\ and\ \bibinfo {author}
  {\bibfnamefont {Y.}~\bibnamefont {Tokura}},\ }\bibfield  {title} {\bibinfo
  {title} {Enhanced directional dichroism of terahertz light in resonance with
  magnetic excitations of the multiferroic
  ${\mathrm{ba}}_{2}{\mathrm{coge}}_{2}{\mathrm{o}}_{7}$ oxide compound},\
  }\href {https://doi.org/10.1103/PhysRevLett.106.057403} {\bibfield  {journal}
  {\bibinfo  {journal} {Phys. Rev. Lett.}\ }\textbf {\bibinfo {volume} {106}},\
  \bibinfo {pages} {057403} (\bibinfo {year} {2011})}\BibitemShut {NoStop}%
\bibitem [{\citenamefont {Takahashi}\ \emph {et~al.}(2012)\citenamefont
  {Takahashi}, \citenamefont {Shimano}, \citenamefont {Kaneko}, \citenamefont
  {Murakawa},\ and\ \citenamefont {Tokura}}]{takahashi2012magnetoelectric}%
  \BibitemOpen
  \bibfield  {author} {\bibinfo {author} {\bibfnamefont {Y.}~\bibnamefont
  {Takahashi}}, \bibinfo {author} {\bibfnamefont {R.}~\bibnamefont {Shimano}},
  \bibinfo {author} {\bibfnamefont {Y.}~\bibnamefont {Kaneko}}, \bibinfo
  {author} {\bibfnamefont {H.}~\bibnamefont {Murakawa}},\ and\ \bibinfo
  {author} {\bibfnamefont {Y.}~\bibnamefont {Tokura}},\ }\bibfield  {title}
  {\bibinfo {title} {Magnetoelectric resonance with electromagnons in a
  perovskite helimagnet},\ }\href
  {https://www.nature.com/articles/nphys2161#citeas} {\bibfield  {journal}
  {\bibinfo  {journal} {Nature Physics}\ }\textbf {\bibinfo {volume} {8}},\
  \bibinfo {pages} {121} (\bibinfo {year} {2012})}\BibitemShut {NoStop}%
\bibitem [{\citenamefont {Kimura}\ \emph {et~al.}(2020)\citenamefont {Kimura},
  \citenamefont {Katsuyoshi}, \citenamefont {Sawada}, \citenamefont {Kimura},\
  and\ \citenamefont {Kimura}}]{kimura2020imaging}%
  \BibitemOpen
  \bibfield  {author} {\bibinfo {author} {\bibfnamefont {K.}~\bibnamefont
  {Kimura}}, \bibinfo {author} {\bibfnamefont {T.}~\bibnamefont {Katsuyoshi}},
  \bibinfo {author} {\bibfnamefont {Y.}~\bibnamefont {Sawada}}, \bibinfo
  {author} {\bibfnamefont {S.}~\bibnamefont {Kimura}},\ and\ \bibinfo {author}
  {\bibfnamefont {T.}~\bibnamefont {Kimura}},\ }\bibfield  {title} {\bibinfo
  {title} {Imaging switchable magnetoelectric quadrupole domains via
  nonreciprocal linear dichroism},\ }\href
  {https://www.nature.com/articles/s43246-020-0040-3#citeas} {\bibfield
  {journal} {\bibinfo  {journal} {Communications Materials}\ }\textbf {\bibinfo
  {volume} {1}},\ \bibinfo {pages} {39} (\bibinfo {year} {2020})}\BibitemShut
  {NoStop}%
\bibitem [{\citenamefont {Sato}\ \emph {et~al.}(2022)\citenamefont {Sato},
  \citenamefont {Abe}, \citenamefont {Tokunaga},\ and\ \citenamefont
  {Arima}}]{PhysRevB.105.094417}%
  \BibitemOpen
  \bibfield  {author} {\bibinfo {author} {\bibfnamefont {T.}~\bibnamefont
  {Sato}}, \bibinfo {author} {\bibfnamefont {N.}~\bibnamefont {Abe}}, \bibinfo
  {author} {\bibfnamefont {Y.}~\bibnamefont {Tokunaga}},\ and\ \bibinfo
  {author} {\bibfnamefont {T.-h.}\ \bibnamefont {Arima}},\ }\bibfield  {title}
  {\bibinfo {title} {Antiferromagnetic domain wall dynamics in magnetoelectric
  ${\mathrm{mntio}}_{3}$ studied by optical imaging},\ }\href
  {https://doi.org/10.1103/PhysRevB.105.094417} {\bibfield  {journal} {\bibinfo
   {journal} {Phys. Rev. B}\ }\textbf {\bibinfo {volume} {105}},\ \bibinfo
  {pages} {094417} (\bibinfo {year} {2022})}\BibitemShut {NoStop}%
\bibitem [{\citenamefont {Hayashida}\ \emph {et~al.}(2022)\citenamefont
  {Hayashida}, \citenamefont {Arakawa}, \citenamefont {Oshima}, \citenamefont
  {Kimura},\ and\ \citenamefont {Kimura}}]{PhysRevResearch.4.043063}%
  \BibitemOpen
  \bibfield  {author} {\bibinfo {author} {\bibfnamefont {T.}~\bibnamefont
  {Hayashida}}, \bibinfo {author} {\bibfnamefont {K.}~\bibnamefont {Arakawa}},
  \bibinfo {author} {\bibfnamefont {T.}~\bibnamefont {Oshima}}, \bibinfo
  {author} {\bibfnamefont {K.}~\bibnamefont {Kimura}},\ and\ \bibinfo {author}
  {\bibfnamefont {T.}~\bibnamefont {Kimura}},\ }\bibfield  {title} {\bibinfo
  {title} {Observation of antiferromagnetic domains in
  ${\mathrm{cr}}_{2}{\mathrm{o}}_{3}$ using nonreciprocal optical effects},\
  }\href {https://doi.org/10.1103/PhysRevResearch.4.043063} {\bibfield
  {journal} {\bibinfo  {journal} {Phys. Rev. Res.}\ }\textbf {\bibinfo {volume}
  {4}},\ \bibinfo {pages} {043063} (\bibinfo {year} {2022})}\BibitemShut
  {NoStop}%
\bibitem [{\citenamefont {Arakawa}\ \emph {et~al.}(2023)\citenamefont
  {Arakawa}, \citenamefont {Hayashida}, \citenamefont {Kimura}, \citenamefont
  {Misawa}, \citenamefont {Nagai}, \citenamefont {Miyamoto}, \citenamefont
  {Okamoto}, \citenamefont {Iga},\ and\ \citenamefont
  {Kimura}}]{PhysRevLett.131.236702}%
  \BibitemOpen
  \bibfield  {author} {\bibinfo {author} {\bibfnamefont {K.}~\bibnamefont
  {Arakawa}}, \bibinfo {author} {\bibfnamefont {T.}~\bibnamefont {Hayashida}},
  \bibinfo {author} {\bibfnamefont {K.}~\bibnamefont {Kimura}}, \bibinfo
  {author} {\bibfnamefont {R.}~\bibnamefont {Misawa}}, \bibinfo {author}
  {\bibfnamefont {T.}~\bibnamefont {Nagai}}, \bibinfo {author} {\bibfnamefont
  {T.}~\bibnamefont {Miyamoto}}, \bibinfo {author} {\bibfnamefont
  {H.}~\bibnamefont {Okamoto}}, \bibinfo {author} {\bibfnamefont
  {F.}~\bibnamefont {Iga}},\ and\ \bibinfo {author} {\bibfnamefont
  {T.}~\bibnamefont {Kimura}},\ }\bibfield  {title} {\bibinfo {title}
  {Detecting magnetoelectric effect in a metallic antiferromagnet via
  nonreciprocal rotation of reflected light},\ }\href
  {https://doi.org/10.1103/PhysRevLett.131.236702} {\bibfield  {journal}
  {\bibinfo  {journal} {Phys. Rev. Lett.}\ }\textbf {\bibinfo {volume} {131}},\
  \bibinfo {pages} {236702} (\bibinfo {year} {2023})}\BibitemShut {NoStop}%
\bibitem [{\citenamefont {Natori}(1975)}]{doi:10.1143/JPSJ.39.1013}%
  \BibitemOpen
  \bibfield  {author} {\bibinfo {author} {\bibfnamefont {K.}~\bibnamefont
  {Natori}},\ }\bibfield  {title} {\bibinfo {title} {Band theory of the optical
  activity of crystals},\ }\href {https://doi.org/10.1143/JPSJ.39.1013}
  {\bibfield  {journal} {\bibinfo  {journal} {Journal of the Physical Society
  of Japan}\ }\textbf {\bibinfo {volume} {39}},\ \bibinfo {pages} {1013}
  (\bibinfo {year} {1975})}\BibitemShut {NoStop}%
\bibitem [{\citenamefont {Zhong}\ \emph {et~al.}(1993)\citenamefont {Zhong},
  \citenamefont {Levine}, \citenamefont {Allan},\ and\ \citenamefont
  {Wilkins}}]{PhysRevB.48.1384}%
  \BibitemOpen
  \bibfield  {author} {\bibinfo {author} {\bibfnamefont {H.}~\bibnamefont
  {Zhong}}, \bibinfo {author} {\bibfnamefont {Z.~H.}\ \bibnamefont {Levine}},
  \bibinfo {author} {\bibfnamefont {D.~C.}\ \bibnamefont {Allan}},\ and\
  \bibinfo {author} {\bibfnamefont {J.~W.}\ \bibnamefont {Wilkins}},\
  }\bibfield  {title} {\bibinfo {title} {Band-theoretic calculations of the
  optical-activity tensor of \ensuremath{\alpha}-quartz and trigonal se},\
  }\href {https://doi.org/10.1103/PhysRevB.48.1384} {\bibfield  {journal}
  {\bibinfo  {journal} {Phys. Rev. B}\ }\textbf {\bibinfo {volume} {48}},\
  \bibinfo {pages} {1384} (\bibinfo {year} {1993})}\BibitemShut {NoStop}%
\bibitem [{\citenamefont {Mineev}\ and\ \citenamefont
  {Yoshioka}(2010)}]{PhysRevB.81.094525}%
  \BibitemOpen
  \bibfield  {author} {\bibinfo {author} {\bibfnamefont {V.~P.}\ \bibnamefont
  {Mineev}}\ and\ \bibinfo {author} {\bibfnamefont {Y.}~\bibnamefont
  {Yoshioka}},\ }\bibfield  {title} {\bibinfo {title} {Optical activity of
  noncentrosymmetric metals},\ }\href
  {https://doi.org/10.1103/PhysRevB.81.094525} {\bibfield  {journal} {\bibinfo
  {journal} {Phys. Rev. B}\ }\textbf {\bibinfo {volume} {81}},\ \bibinfo
  {pages} {094525} (\bibinfo {year} {2010})}\BibitemShut {NoStop}%
\bibitem [{\citenamefont {Malashevich}\ and\ \citenamefont
  {Souza}(2010)}]{PhysRevB.82.245118}%
  \BibitemOpen
  \bibfield  {author} {\bibinfo {author} {\bibfnamefont {A.}~\bibnamefont
  {Malashevich}}\ and\ \bibinfo {author} {\bibfnamefont {I.}~\bibnamefont
  {Souza}},\ }\bibfield  {title} {\bibinfo {title} {Band theory of spatial
  dispersion in magnetoelectrics},\ }\href
  {https://doi.org/10.1103/PhysRevB.82.245118} {\bibfield  {journal} {\bibinfo
  {journal} {Phys. Rev. B}\ }\textbf {\bibinfo {volume} {82}},\ \bibinfo
  {pages} {245118} (\bibinfo {year} {2010})}\BibitemShut {NoStop}%
\bibitem [{\citenamefont {Mineev}(2013)}]{PhysRevB.88.134514}%
  \BibitemOpen
  \bibfield  {author} {\bibinfo {author} {\bibfnamefont {V.~P.}\ \bibnamefont
  {Mineev}},\ }\bibfield  {title} {\bibinfo {title} {Magnetostatics and optics
  of noncentrosymmetric metals},\ }\href
  {https://doi.org/10.1103/PhysRevB.88.134514} {\bibfield  {journal} {\bibinfo
  {journal} {Phys. Rev. B}\ }\textbf {\bibinfo {volume} {88}},\ \bibinfo
  {pages} {134514} (\bibinfo {year} {2013})}\BibitemShut {NoStop}%
\bibitem [{\citenamefont {Zhong}\ \emph {et~al.}(2016)\citenamefont {Zhong},
  \citenamefont {Moore},\ and\ \citenamefont {Souza}}]{PhysRevLett.116.077201}%
  \BibitemOpen
  \bibfield  {author} {\bibinfo {author} {\bibfnamefont {S.}~\bibnamefont
  {Zhong}}, \bibinfo {author} {\bibfnamefont {J.~E.}\ \bibnamefont {Moore}},\
  and\ \bibinfo {author} {\bibfnamefont {I.}~\bibnamefont {Souza}},\ }\bibfield
   {title} {\bibinfo {title} {Gyrotropic magnetic effect and the magnetic
  moment on the fermi surface},\ }\href
  {https://doi.org/10.1103/PhysRevLett.116.077201} {\bibfield  {journal}
  {\bibinfo  {journal} {Phys. Rev. Lett.}\ }\textbf {\bibinfo {volume} {116}},\
  \bibinfo {pages} {077201} (\bibinfo {year} {2016})}\BibitemShut {NoStop}%
\bibitem [{\citenamefont {Ma}\ and\ \citenamefont
  {Pesin}(2015)}]{PhysRevB.92.235205}%
  \BibitemOpen
  \bibfield  {author} {\bibinfo {author} {\bibfnamefont {J.}~\bibnamefont
  {Ma}}\ and\ \bibinfo {author} {\bibfnamefont {D.~A.}\ \bibnamefont {Pesin}},\
  }\bibfield  {title} {\bibinfo {title} {Chiral magnetic effect and natural
  optical activity in metals with or without weyl points},\ }\href
  {https://doi.org/10.1103/PhysRevB.92.235205} {\bibfield  {journal} {\bibinfo
  {journal} {Phys. Rev. B}\ }\textbf {\bibinfo {volume} {92}},\ \bibinfo
  {pages} {235205} (\bibinfo {year} {2015})}\BibitemShut {NoStop}%
\bibitem [{\citenamefont {Gao}\ and\ \citenamefont
  {Xiao}(2019)}]{PhysRevLett.122.227402}%
  \BibitemOpen
  \bibfield  {author} {\bibinfo {author} {\bibfnamefont {Y.}~\bibnamefont
  {Gao}}\ and\ \bibinfo {author} {\bibfnamefont {D.}~\bibnamefont {Xiao}},\
  }\bibfield  {title} {\bibinfo {title} {Nonreciprocal directional dichroism
  induced by the quantum metric dipole},\ }\href
  {https://doi.org/10.1103/PhysRevLett.122.227402} {\bibfield  {journal}
  {\bibinfo  {journal} {Phys. Rev. Lett.}\ }\textbf {\bibinfo {volume} {122}},\
  \bibinfo {pages} {227402} (\bibinfo {year} {2019})}\BibitemShut {NoStop}%
\bibitem [{\citenamefont {Duff}\ and\ \citenamefont
  {Sipe}(2022)}]{PhysRevB.106.085413}%
  \BibitemOpen
  \bibfield  {author} {\bibinfo {author} {\bibfnamefont {A.~H.}\ \bibnamefont
  {Duff}}\ and\ \bibinfo {author} {\bibfnamefont {J.~E.}\ \bibnamefont
  {Sipe}},\ }\bibfield  {title} {\bibinfo {title} {Magnetoelectric
  polarizability and optical activity: Spin and frequency dependence},\ }\href
  {https://doi.org/10.1103/PhysRevB.106.085413} {\bibfield  {journal} {\bibinfo
   {journal} {Phys. Rev. B}\ }\textbf {\bibinfo {volume} {106}},\ \bibinfo
  {pages} {085413} (\bibinfo {year} {2022})}\BibitemShut {NoStop}%
\bibitem [{\citenamefont {Óscar Pozo~Ocaña}\ and\ \citenamefont
  {Souza}(2023)}]{10.21468/SciPostPhys.14.5.118}%
  \BibitemOpen
  \bibfield  {author} {\bibinfo {author} {\bibnamefont {Óscar Pozo~Ocaña}}\
  and\ \bibinfo {author} {\bibfnamefont {I.}~\bibnamefont {Souza}},\ }\bibfield
   {title} {\bibinfo {title} {{Multipole theory of optical spatial dispersion
  in crystals}},\ }\href {https://doi.org/10.21468/SciPostPhys.14.5.118}
  {\bibfield  {journal} {\bibinfo  {journal} {SciPost Phys.}\ }\textbf
  {\bibinfo {volume} {14}},\ \bibinfo {pages} {118} (\bibinfo {year}
  {2023})}\BibitemShut {NoStop}%
\bibitem [{\citenamefont {Hidalgo}\ \emph {et~al.}(2009)\citenamefont
  {Hidalgo}, \citenamefont {S\'anchez-Castillo},\ and\ \citenamefont
  {Noguez}}]{PhysRevB.79.075438}%
  \BibitemOpen
  \bibfield  {author} {\bibinfo {author} {\bibfnamefont {F.}~\bibnamefont
  {Hidalgo}}, \bibinfo {author} {\bibfnamefont {A.}~\bibnamefont
  {S\'anchez-Castillo}},\ and\ \bibinfo {author} {\bibfnamefont
  {C.}~\bibnamefont {Noguez}},\ }\bibfield  {title} {\bibinfo {title}
  {Efficient first-principles method for calculating the circular dichroism of
  nanostructures},\ }\href {https://doi.org/10.1103/PhysRevB.79.075438}
  {\bibfield  {journal} {\bibinfo  {journal} {Phys. Rev. B}\ }\textbf {\bibinfo
  {volume} {79}},\ \bibinfo {pages} {075438} (\bibinfo {year}
  {2009})}\BibitemShut {NoStop}%
\bibitem [{\citenamefont {Tsirkin}\ \emph {et~al.}(2018)\citenamefont
  {Tsirkin}, \citenamefont {Puente},\ and\ \citenamefont
  {Souza}}]{PhysRevB.97.035158}%
  \BibitemOpen
  \bibfield  {author} {\bibinfo {author} {\bibfnamefont {S.~S.}\ \bibnamefont
  {Tsirkin}}, \bibinfo {author} {\bibfnamefont {P.~A.}\ \bibnamefont
  {Puente}},\ and\ \bibinfo {author} {\bibfnamefont {I.}~\bibnamefont
  {Souza}},\ }\bibfield  {title} {\bibinfo {title} {Gyrotropic effects in
  trigonal tellurium studied from first principles},\ }\href
  {https://doi.org/10.1103/PhysRevB.97.035158} {\bibfield  {journal} {\bibinfo
  {journal} {Phys. Rev. B}\ }\textbf {\bibinfo {volume} {97}},\ \bibinfo
  {pages} {035158} (\bibinfo {year} {2018})}\BibitemShut {NoStop}%
\bibitem [{\citenamefont {R{\'e}rat}\ and\ \citenamefont
  {Kirtman}(2021)}]{rerat2021first}%
  \BibitemOpen
  \bibfield  {author} {\bibinfo {author} {\bibfnamefont {M.}~\bibnamefont
  {R{\'e}rat}}\ and\ \bibinfo {author} {\bibfnamefont {B.}~\bibnamefont
  {Kirtman}},\ }\bibfield  {title} {\bibinfo {title} {First-principles
  calculation of the optical rotatory power of periodic systems: Application on
  $\alpha$-quartz, tartaric acid crystal, and chiral (n, m)-carbon nanotubes},\
  }\href {https://pubs.acs.org/doi/10.1021/acs.jctc.1c00243} {\bibfield
  {journal} {\bibinfo  {journal} {Journal of Chemical Theory and Computation}\
  }\textbf {\bibinfo {volume} {17}},\ \bibinfo {pages} {4063} (\bibinfo {year}
  {2021})}\BibitemShut {NoStop}%
\bibitem [{\citenamefont {Wang}\ and\ \citenamefont
  {Yan}(2023)}]{PhysRevB.107.045201}%
  \BibitemOpen
  \bibfield  {author} {\bibinfo {author} {\bibfnamefont {X.}~\bibnamefont
  {Wang}}\ and\ \bibinfo {author} {\bibfnamefont {Y.}~\bibnamefont {Yan}},\
  }\bibfield  {title} {\bibinfo {title} {Optical activity of solids from first
  principles},\ }\href {https://doi.org/10.1103/PhysRevB.107.045201} {\bibfield
   {journal} {\bibinfo  {journal} {Phys. Rev. B}\ }\textbf {\bibinfo {volume}
  {107}},\ \bibinfo {pages} {045201} (\bibinfo {year} {2023})}\BibitemShut
  {NoStop}%
\bibitem [{\citenamefont {Morell}\ \emph {et~al.}(2017)\citenamefont {Morell},
  \citenamefont {Chico},\ and\ \citenamefont {Brey}}]{morell2017twisting}%
  \BibitemOpen
  \bibfield  {author} {\bibinfo {author} {\bibfnamefont {E.~S.}\ \bibnamefont
  {Morell}}, \bibinfo {author} {\bibfnamefont {L.}~\bibnamefont {Chico}},\ and\
  \bibinfo {author} {\bibfnamefont {L.}~\bibnamefont {Brey}},\ }\bibfield
  {title} {\bibinfo {title} {Twisting dirac fermions: circular dichroism in
  bilayer graphene},\ }\href
  {https://iopscience.iop.org/article/10.1088/2053-1583/aa7eb6} {\bibfield
  {journal} {\bibinfo  {journal} {2D Materials}\ }\textbf {\bibinfo {volume}
  {4}},\ \bibinfo {pages} {035015} (\bibinfo {year} {2017})}\BibitemShut
  {NoStop}%
\bibitem [{\citenamefont {Stauber}\ \emph {et~al.}(2018)\citenamefont
  {Stauber}, \citenamefont {Low},\ and\ \citenamefont
  {G\'omez-Santos}}]{PhysRevLett.120.046801}%
  \BibitemOpen
  \bibfield  {author} {\bibinfo {author} {\bibfnamefont {T.}~\bibnamefont
  {Stauber}}, \bibinfo {author} {\bibfnamefont {T.}~\bibnamefont {Low}},\ and\
  \bibinfo {author} {\bibfnamefont {G.}~\bibnamefont {G\'omez-Santos}},\
  }\bibfield  {title} {\bibinfo {title} {Chiral response of twisted bilayer
  graphene},\ }\href {https://doi.org/10.1103/PhysRevLett.120.046801}
  {\bibfield  {journal} {\bibinfo  {journal} {Phys. Rev. Lett.}\ }\textbf
  {\bibinfo {volume} {120}},\ \bibinfo {pages} {046801} (\bibinfo {year}
  {2018})}\BibitemShut {NoStop}%
\bibitem [{\citenamefont {Chang}\ \emph {et~al.}(2022)\citenamefont {Chang},
  \citenamefont {Zheng}, \citenamefont {Sipe},\ and\ \citenamefont
  {Cheng}}]{PhysRevB.106.245405}%
  \BibitemOpen
  \bibfield  {author} {\bibinfo {author} {\bibfnamefont {K.}~\bibnamefont
  {Chang}}, \bibinfo {author} {\bibfnamefont {Z.}~\bibnamefont {Zheng}},
  \bibinfo {author} {\bibfnamefont {J.~E.}\ \bibnamefont {Sipe}},\ and\
  \bibinfo {author} {\bibfnamefont {J.~L.}\ \bibnamefont {Cheng}},\ }\bibfield
  {title} {\bibinfo {title} {Theory of optical activity in doped systems with
  application to twisted bilayer graphene},\ }\href
  {https://doi.org/10.1103/PhysRevB.106.245405} {\bibfield  {journal} {\bibinfo
   {journal} {Phys. Rev. B}\ }\textbf {\bibinfo {volume} {106}},\ \bibinfo
  {pages} {245405} (\bibinfo {year} {2022})}\BibitemShut {NoStop}%
\bibitem [{\citenamefont {Ho}\ and\ \citenamefont
  {Do}(2023)}]{PhysRevB.107.195141}%
  \BibitemOpen
  \bibfield  {author} {\bibinfo {author} {\bibfnamefont {S.~T.}\ \bibnamefont
  {Ho}}\ and\ \bibinfo {author} {\bibfnamefont {V.~N.}\ \bibnamefont {Do}},\
  }\bibfield  {title} {\bibinfo {title} {Optical activity and transport in
  twisted bilayer graphene: Spatial dispersion effects},\ }\href
  {https://doi.org/10.1103/PhysRevB.107.195141} {\bibfield  {journal} {\bibinfo
   {journal} {Phys. Rev. B}\ }\textbf {\bibinfo {volume} {107}},\ \bibinfo
  {pages} {195141} (\bibinfo {year} {2023})}\BibitemShut {NoStop}%
\bibitem [{\citenamefont {Ahn}\ \emph {et~al.}(2022)\citenamefont {Ahn},
  \citenamefont {Xu},\ and\ \citenamefont {Vishwanath}}]{ahn2022theory}%
  \BibitemOpen
  \bibfield  {author} {\bibinfo {author} {\bibfnamefont {J.}~\bibnamefont
  {Ahn}}, \bibinfo {author} {\bibfnamefont {S.-Y.}\ \bibnamefont {Xu}},\ and\
  \bibinfo {author} {\bibfnamefont {A.}~\bibnamefont {Vishwanath}},\ }\bibfield
   {title} {\bibinfo {title} {Theory of optical axion electrodynamics and
  application to the kerr effect in topological antiferromagnets},\ }\href
  {https://www.nature.com/articles/s41467-022-35248-8} {\bibfield  {journal}
  {\bibinfo  {journal} {Nature Communications}\ }\textbf {\bibinfo {volume}
  {13}},\ \bibinfo {pages} {7615} (\bibinfo {year} {2022})}\BibitemShut
  {NoStop}%
\bibitem [{\citenamefont {Sekh}\ and\ \citenamefont
  {Mandal}(2022)}]{PhysRevB.105.235403}%
  \BibitemOpen
  \bibfield  {author} {\bibinfo {author} {\bibfnamefont {S.}~\bibnamefont
  {Sekh}}\ and\ \bibinfo {author} {\bibfnamefont {I.}~\bibnamefont {Mandal}},\
  }\bibfield  {title} {\bibinfo {title} {Circular dichroism as a probe for
  topology in three-dimensional semimetals},\ }\href
  {https://doi.org/10.1103/PhysRevB.105.235403} {\bibfield  {journal} {\bibinfo
   {journal} {Phys. Rev. B}\ }\textbf {\bibinfo {volume} {105}},\ \bibinfo
  {pages} {235403} (\bibinfo {year} {2022})}\BibitemShut {NoStop}%
\bibitem [{\citenamefont {Mandal}(2023)}]{mandal2023signatures}%
  \BibitemOpen
  \bibfield  {author} {\bibinfo {author} {\bibfnamefont {I.}~\bibnamefont
  {Mandal}},\ }\bibfield  {title} {\bibinfo {title} {Signatures of two-and
  three-dimensional semimetals from circular dichroism},\ }\href
  {https://www.worldscientific.com/doi/10.1142/S0217979224502163} {\bibfield
  {journal} {\bibinfo  {journal} {International Journal of Modern Physics B}\
  ,\ \bibinfo {pages} {2450216}} (\bibinfo {year} {2023})}\BibitemShut
  {NoStop}%
\bibitem [{\citenamefont {Ahn}\ and\ \citenamefont
  {Ghosh}(2023)}]{PhysRevLett.131.116603}%
  \BibitemOpen
  \bibfield  {author} {\bibinfo {author} {\bibfnamefont {J.}~\bibnamefont
  {Ahn}}\ and\ \bibinfo {author} {\bibfnamefont {B.}~\bibnamefont {Ghosh}},\
  }\bibfield  {title} {\bibinfo {title} {Topological circular dichroism in
  chiral multifold semimetals},\ }\href
  {https://doi.org/10.1103/PhysRevLett.131.116603} {\bibfield  {journal}
  {\bibinfo  {journal} {Phys. Rev. Lett.}\ }\textbf {\bibinfo {volume} {131}},\
  \bibinfo {pages} {116603} (\bibinfo {year} {2023})}\BibitemShut {NoStop}%
\bibitem [{\citenamefont {Shinada}\ and\ \citenamefont
  {Peters}(2023{\natexlab{a}})}]{PhysRevB.108.165119}%
  \BibitemOpen
  \bibfield  {author} {\bibinfo {author} {\bibfnamefont {K.}~\bibnamefont
  {Shinada}}\ and\ \bibinfo {author} {\bibfnamefont {R.}~\bibnamefont
  {Peters}},\ }\bibfield  {title} {\bibinfo {title} {Unique properties of the
  optical activity in noncentrosymmetric superconductors: Sum rule, missing
  area, and relation with the superconducting edelstein effect},\ }\href
  {https://doi.org/10.1103/PhysRevB.108.165119} {\bibfield  {journal} {\bibinfo
   {journal} {Phys. Rev. B}\ }\textbf {\bibinfo {volume} {108}},\ \bibinfo
  {pages} {165119} (\bibinfo {year} {2023}{\natexlab{a}})}\BibitemShut
  {NoStop}%
\bibitem [{\citenamefont {Shinada}\ \emph {et~al.}(2023)\citenamefont
  {Shinada}, \citenamefont {Kofuji},\ and\ \citenamefont
  {Peters}}]{PhysRevB.107.094106}%
  \BibitemOpen
  \bibfield  {author} {\bibinfo {author} {\bibfnamefont {K.}~\bibnamefont
  {Shinada}}, \bibinfo {author} {\bibfnamefont {A.}~\bibnamefont {Kofuji}},\
  and\ \bibinfo {author} {\bibfnamefont {R.}~\bibnamefont {Peters}},\
  }\bibfield  {title} {\bibinfo {title} {Quantum theory of the intrinsic
  orbital magnetoelectric effect in itinerant electron systems at finite
  temperatures},\ }\href {https://doi.org/10.1103/PhysRevB.107.094106}
  {\bibfield  {journal} {\bibinfo  {journal} {Phys. Rev. B}\ }\textbf {\bibinfo
  {volume} {107}},\ \bibinfo {pages} {094106} (\bibinfo {year}
  {2023})}\BibitemShut {NoStop}%
\bibitem [{\citenamefont {Shinada}\ and\ \citenamefont
  {Peters}(2023{\natexlab{b}})}]{PhysRevB.107.214109}%
  \BibitemOpen
  \bibfield  {author} {\bibinfo {author} {\bibfnamefont {K.}~\bibnamefont
  {Shinada}}\ and\ \bibinfo {author} {\bibfnamefont {R.}~\bibnamefont
  {Peters}},\ }\bibfield  {title} {\bibinfo {title} {Orbital
  gravitomagnetoelectric response and orbital magnetic quadrupole moment
  correction},\ }\href {https://doi.org/10.1103/PhysRevB.107.214109} {\bibfield
   {journal} {\bibinfo  {journal} {Phys. Rev. B}\ }\textbf {\bibinfo {volume}
  {107}},\ \bibinfo {pages} {214109} (\bibinfo {year}
  {2023}{\natexlab{b}})}\BibitemShut {NoStop}%
\bibitem [{\citenamefont {Fukushima}\ \emph {et~al.}(2008)\citenamefont
  {Fukushima}, \citenamefont {Kharzeev},\ and\ \citenamefont
  {Warringa}}]{PhysRevD.78.074033}%
  \BibitemOpen
  \bibfield  {author} {\bibinfo {author} {\bibfnamefont {K.}~\bibnamefont
  {Fukushima}}, \bibinfo {author} {\bibfnamefont {D.~E.}\ \bibnamefont
  {Kharzeev}},\ and\ \bibinfo {author} {\bibfnamefont {H.~J.}\ \bibnamefont
  {Warringa}},\ }\bibfield  {title} {\bibinfo {title} {Chiral magnetic
  effect},\ }\href {https://doi.org/10.1103/PhysRevD.78.074033} {\bibfield
  {journal} {\bibinfo  {journal} {Phys. Rev. D}\ }\textbf {\bibinfo {volume}
  {78}},\ \bibinfo {pages} {074033} (\bibinfo {year} {2008})}\BibitemShut
  {NoStop}%
\bibitem [{\citenamefont {Tinkham}(2004)}]{tinkham2004superconductivity}%
  \BibitemOpen
  \bibfield  {author} {\bibinfo {author} {\bibfnamefont {M.}~\bibnamefont
  {Tinkham}},\ }\href@noop {} {\emph {\bibinfo {title} {Introduction to
  Superconductivity}}}\ (\bibinfo  {publisher} {Courier Corporation},\ \bibinfo
  {year} {2004})\BibitemShut {NoStop}%
\bibitem [{\citenamefont {Watanabe}\ \emph
  {et~al.}(2022{\natexlab{a}})\citenamefont {Watanabe}, \citenamefont {Daido},\
  and\ \citenamefont {Yanase}}]{PhysRevB.105.024308}%
  \BibitemOpen
  \bibfield  {author} {\bibinfo {author} {\bibfnamefont {H.}~\bibnamefont
  {Watanabe}}, \bibinfo {author} {\bibfnamefont {A.}~\bibnamefont {Daido}},\
  and\ \bibinfo {author} {\bibfnamefont {Y.}~\bibnamefont {Yanase}},\
  }\bibfield  {title} {\bibinfo {title} {Nonreciprocal optical response in
  parity-breaking superconductors},\ }\href
  {https://doi.org/10.1103/PhysRevB.105.024308} {\bibfield  {journal} {\bibinfo
   {journal} {Phys. Rev. B}\ }\textbf {\bibinfo {volume} {105}},\ \bibinfo
  {pages} {024308} (\bibinfo {year} {2022}{\natexlab{a}})}\BibitemShut
  {NoStop}%
\bibitem [{\citenamefont {Watanabe}\ \emph
  {et~al.}(2022{\natexlab{b}})\citenamefont {Watanabe}, \citenamefont {Daido},\
  and\ \citenamefont {Yanase}}]{PhysRevB.105.L100504}%
  \BibitemOpen
  \bibfield  {author} {\bibinfo {author} {\bibfnamefont {H.}~\bibnamefont
  {Watanabe}}, \bibinfo {author} {\bibfnamefont {A.}~\bibnamefont {Daido}},\
  and\ \bibinfo {author} {\bibfnamefont {Y.}~\bibnamefont {Yanase}},\
  }\bibfield  {title} {\bibinfo {title} {Nonreciprocal meissner response in
  parity-mixed superconductors},\ }\href
  {https://doi.org/10.1103/PhysRevB.105.L100504} {\bibfield  {journal}
  {\bibinfo  {journal} {Phys. Rev. B}\ }\textbf {\bibinfo {volume} {105}},\
  \bibinfo {pages} {L100504} (\bibinfo {year}
  {2022}{\natexlab{b}})}\BibitemShut {NoStop}%
\bibitem [{\citenamefont {Hornreich}\ and\ \citenamefont
  {Shtrikman}(1968)}]{PhysRev.171.1065}%
  \BibitemOpen
  \bibfield  {author} {\bibinfo {author} {\bibfnamefont {R.~M.}\ \bibnamefont
  {Hornreich}}\ and\ \bibinfo {author} {\bibfnamefont {S.}~\bibnamefont
  {Shtrikman}},\ }\bibfield  {title} {\bibinfo {title} {Theory of gyrotropic
  birefringence},\ }\href {https://doi.org/10.1103/PhysRev.171.1065} {\bibfield
   {journal} {\bibinfo  {journal} {Phys. Rev.}\ }\textbf {\bibinfo {volume}
  {171}},\ \bibinfo {pages} {1065} (\bibinfo {year} {1968})}\BibitemShut
  {NoStop}%
\bibitem [{\citenamefont {Levitov}\ \emph {et~al.}(1985)\citenamefont
  {Levitov}, \citenamefont {Nazarov},\ and\ \citenamefont
  {Eliashberg}}]{levitov1985magnetoelectric}%
  \BibitemOpen
  \bibfield  {author} {\bibinfo {author} {\bibfnamefont {L.~S.}\ \bibnamefont
  {Levitov}}, \bibinfo {author} {\bibfnamefont {Y.~V.}\ \bibnamefont
  {Nazarov}},\ and\ \bibinfo {author} {\bibfnamefont {G.~M.}\ \bibnamefont
  {Eliashberg}},\ }\bibfield  {title} {\bibinfo {title} {Magnetoelectric
  effects in conductors with mirror isomer symmetry},\ }\href@noop {}
  {\bibfield  {journal} {\bibinfo  {journal} {Soviet Physics JETP}\ }\textbf
  {\bibinfo {volume} {61}},\ \bibinfo {pages} {133} (\bibinfo {year}
  {1985})}\BibitemShut {NoStop}%
\bibitem [{\citenamefont {Edelstein}(1995)}]{PhysRevLett.75.2004}%
  \BibitemOpen
  \bibfield  {author} {\bibinfo {author} {\bibfnamefont {V.~M.}\ \bibnamefont
  {Edelstein}},\ }\bibfield  {title} {\bibinfo {title} {Magnetoelectric effect
  in polar superconductors},\ }\href
  {https://doi.org/10.1103/PhysRevLett.75.2004} {\bibfield  {journal} {\bibinfo
   {journal} {Phys. Rev. Lett.}\ }\textbf {\bibinfo {volume} {75}},\ \bibinfo
  {pages} {2004} (\bibinfo {year} {1995})}\BibitemShut {NoStop}%
\bibitem [{\citenamefont {Gor'kov}\ and\ \citenamefont
  {Rashba}(2001)}]{PhysRevLett.87.037004}%
  \BibitemOpen
  \bibfield  {author} {\bibinfo {author} {\bibfnamefont {L.~P.}\ \bibnamefont
  {Gor'kov}}\ and\ \bibinfo {author} {\bibfnamefont {E.~I.}\ \bibnamefont
  {Rashba}},\ }\bibfield  {title} {\bibinfo {title} {Superconducting 2d system
  with lifted spin degeneracy: Mixed singlet-triplet state},\ }\href
  {https://doi.org/10.1103/PhysRevLett.87.037004} {\bibfield  {journal}
  {\bibinfo  {journal} {Phys. Rev. Lett.}\ }\textbf {\bibinfo {volume} {87}},\
  \bibinfo {pages} {037004} (\bibinfo {year} {2001})}\BibitemShut {NoStop}%
\bibitem [{\citenamefont {Yip}(2002)}]{PhysRevB.65.144508}%
  \BibitemOpen
  \bibfield  {author} {\bibinfo {author} {\bibfnamefont {S.~K.}\ \bibnamefont
  {Yip}},\ }\bibfield  {title} {\bibinfo {title} {Two-dimensional
  superconductivity with strong spin-orbit interaction},\ }\href
  {https://doi.org/10.1103/PhysRevB.65.144508} {\bibfield  {journal} {\bibinfo
  {journal} {Phys. Rev. B}\ }\textbf {\bibinfo {volume} {65}},\ \bibinfo
  {pages} {144508} (\bibinfo {year} {2002})}\BibitemShut {NoStop}%
\bibitem [{\citenamefont {Fujimoto}(2005)}]{PhysRevB.72.024515}%
  \BibitemOpen
  \bibfield  {author} {\bibinfo {author} {\bibfnamefont {S.}~\bibnamefont
  {Fujimoto}},\ }\bibfield  {title} {\bibinfo {title} {Magnetoelectric effects
  in heavy-fermion superconductors without inversion symmetry},\ }\href
  {https://doi.org/10.1103/PhysRevB.72.024515} {\bibfield  {journal} {\bibinfo
  {journal} {Phys. Rev. B}\ }\textbf {\bibinfo {volume} {72}},\ \bibinfo
  {pages} {024515} (\bibinfo {year} {2005})}\BibitemShut {NoStop}%
\bibitem [{\citenamefont {Lu}\ and\ \citenamefont
  {Yip}(2008)}]{PhysRevB.77.054515}%
  \BibitemOpen
  \bibfield  {author} {\bibinfo {author} {\bibfnamefont {C.-K.}\ \bibnamefont
  {Lu}}\ and\ \bibinfo {author} {\bibfnamefont {S.}~\bibnamefont {Yip}},\
  }\bibfield  {title} {\bibinfo {title} {Signature of superconducting states in
  cubic crystal without inversion symmetry},\ }\href
  {https://doi.org/10.1103/PhysRevB.77.054515} {\bibfield  {journal} {\bibinfo
  {journal} {Phys. Rev. B}\ }\textbf {\bibinfo {volume} {77}},\ \bibinfo
  {pages} {054515} (\bibinfo {year} {2008})}\BibitemShut {NoStop}%
\bibitem [{\citenamefont {He}\ and\ \citenamefont
  {Law}(2021)}]{PhysRevResearch.3.L032012}%
  \BibitemOpen
  \bibfield  {author} {\bibinfo {author} {\bibfnamefont {W.-Y.}\ \bibnamefont
  {He}}\ and\ \bibinfo {author} {\bibfnamefont {K.~T.}\ \bibnamefont {Law}},\
  }\bibfield  {title} {\bibinfo {title} {Superconducting orbital
  magnetoelectric effect and its evolution across the superconductor-normal
  metal phase transition},\ }\href
  {https://doi.org/10.1103/PhysRevResearch.3.L032012} {\bibfield  {journal}
  {\bibinfo  {journal} {Phys. Rev. Res.}\ }\textbf {\bibinfo {volume} {3}},\
  \bibinfo {pages} {L032012} (\bibinfo {year} {2021})}\BibitemShut {NoStop}%
\bibitem [{\citenamefont {Cheiwchanchamnangij}\ and\ \citenamefont
  {Lambrecht}(2012)}]{PhysRevB.85.205302}%
  \BibitemOpen
  \bibfield  {author} {\bibinfo {author} {\bibfnamefont {T.}~\bibnamefont
  {Cheiwchanchamnangij}}\ and\ \bibinfo {author} {\bibfnamefont {W.~R.~L.}\
  \bibnamefont {Lambrecht}},\ }\bibfield  {title} {\bibinfo {title}
  {Quasiparticle band structure calculation of monolayer, bilayer, and bulk
  mos${}_{2}$},\ }\href {https://doi.org/10.1103/PhysRevB.85.205302} {\bibfield
   {journal} {\bibinfo  {journal} {Phys. Rev. B}\ }\textbf {\bibinfo {volume}
  {85}},\ \bibinfo {pages} {205302} (\bibinfo {year} {2012})}\BibitemShut
  {NoStop}%
\bibitem [{\citenamefont {Kadantsev}\ and\ \citenamefont
  {Hawrylak}(2012)}]{KADANTSEV2012909}%
  \BibitemOpen
  \bibfield  {author} {\bibinfo {author} {\bibfnamefont {E.~S.}\ \bibnamefont
  {Kadantsev}}\ and\ \bibinfo {author} {\bibfnamefont {P.}~\bibnamefont
  {Hawrylak}},\ }\bibfield  {title} {\bibinfo {title} {Electronic structure of
  a single mos2 monolayer},\ }\href
  {https://doi.org/https://doi.org/10.1016/j.ssc.2012.02.005} {\bibfield
  {journal} {\bibinfo  {journal} {Solid State Communications}\ }\textbf
  {\bibinfo {volume} {152}},\ \bibinfo {pages} {909} (\bibinfo {year}
  {2012})}\BibitemShut {NoStop}%
\bibitem [{\citenamefont {He}\ and\ \citenamefont
  {Law}(2020)}]{PhysRevResearch.2.012073}%
  \BibitemOpen
  \bibfield  {author} {\bibinfo {author} {\bibfnamefont {W.-Y.}\ \bibnamefont
  {He}}\ and\ \bibinfo {author} {\bibfnamefont {K.~T.}\ \bibnamefont {Law}},\
  }\bibfield  {title} {\bibinfo {title} {Magnetoelectric effects in gyrotropic
  superconductors},\ }\href {https://doi.org/10.1103/PhysRevResearch.2.012073}
  {\bibfield  {journal} {\bibinfo  {journal} {Phys. Rev. Res.}\ }\textbf
  {\bibinfo {volume} {2}},\ \bibinfo {pages} {012073} (\bibinfo {year}
  {2020})}\BibitemShut {NoStop}%
\bibitem [{\citenamefont {Kim}\ \emph {et~al.}(2016)\citenamefont {Kim},
  \citenamefont {S{\'a}nchez-Castillo}, \citenamefont {Ziegler}, \citenamefont
  {Ogawa}, \citenamefont {Noguez},\ and\ \citenamefont {Park}}]{kim2016chiral}%
  \BibitemOpen
  \bibfield  {author} {\bibinfo {author} {\bibfnamefont {C.-J.}\ \bibnamefont
  {Kim}}, \bibinfo {author} {\bibfnamefont {A.}~\bibnamefont
  {S{\'a}nchez-Castillo}}, \bibinfo {author} {\bibfnamefont {Z.}~\bibnamefont
  {Ziegler}}, \bibinfo {author} {\bibfnamefont {Y.}~\bibnamefont {Ogawa}},
  \bibinfo {author} {\bibfnamefont {C.}~\bibnamefont {Noguez}},\ and\ \bibinfo
  {author} {\bibfnamefont {J.}~\bibnamefont {Park}},\ }\bibfield  {title}
  {\bibinfo {title} {Chiral atomically thin films},\ }\href
  {https://www.nature.com/articles/nnano.2016.3#Sec6} {\bibfield  {journal}
  {\bibinfo  {journal} {Nature nanotechnology}\ }\textbf {\bibinfo {volume}
  {11}},\ \bibinfo {pages} {520} (\bibinfo {year} {2016})}\BibitemShut
  {NoStop}%
\bibitem [{\citenamefont {Xiao}\ \emph {et~al.}(2007)\citenamefont {Xiao},
  \citenamefont {Yao},\ and\ \citenamefont {Niu}}]{PhysRevLett.99.236809}%
  \BibitemOpen
  \bibfield  {author} {\bibinfo {author} {\bibfnamefont {D.}~\bibnamefont
  {Xiao}}, \bibinfo {author} {\bibfnamefont {W.}~\bibnamefont {Yao}},\ and\
  \bibinfo {author} {\bibfnamefont {Q.}~\bibnamefont {Niu}},\ }\bibfield
  {title} {\bibinfo {title} {Valley-contrasting physics in graphene: Magnetic
  moment and topological transport},\ }\href
  {https://doi.org/10.1103/PhysRevLett.99.236809} {\bibfield  {journal}
  {\bibinfo  {journal} {Phys. Rev. Lett.}\ }\textbf {\bibinfo {volume} {99}},\
  \bibinfo {pages} {236809} (\bibinfo {year} {2007})}\BibitemShut {NoStop}%
\bibitem [{\citenamefont {Sodemann}\ and\ \citenamefont
  {Fu}(2015)}]{PhysRevLett.115.216806}%
  \BibitemOpen
  \bibfield  {author} {\bibinfo {author} {\bibfnamefont {I.}~\bibnamefont
  {Sodemann}}\ and\ \bibinfo {author} {\bibfnamefont {L.}~\bibnamefont {Fu}},\
  }\bibfield  {title} {\bibinfo {title} {Quantum nonlinear hall effect induced
  by berry curvature dipole in time-reversal invariant materials},\ }\href
  {https://doi.org/10.1103/PhysRevLett.115.216806} {\bibfield  {journal}
  {\bibinfo  {journal} {Phys. Rev. Lett.}\ }\textbf {\bibinfo {volume} {115}},\
  \bibinfo {pages} {216806} (\bibinfo {year} {2015})}\BibitemShut {NoStop}%
\bibitem [{\citenamefont {Son}\ \emph {et~al.}(2019)\citenamefont {Son},
  \citenamefont {Kim}, \citenamefont {Ahn}, \citenamefont {Lee},\ and\
  \citenamefont {Lee}}]{PhysRevLett.123.036806}%
  \BibitemOpen
  \bibfield  {author} {\bibinfo {author} {\bibfnamefont {J.}~\bibnamefont
  {Son}}, \bibinfo {author} {\bibfnamefont {K.-H.}\ \bibnamefont {Kim}},
  \bibinfo {author} {\bibfnamefont {Y.~H.}\ \bibnamefont {Ahn}}, \bibinfo
  {author} {\bibfnamefont {H.-W.}\ \bibnamefont {Lee}},\ and\ \bibinfo {author}
  {\bibfnamefont {J.}~\bibnamefont {Lee}},\ }\bibfield  {title} {\bibinfo
  {title} {Strain engineering of the berry curvature dipole and valley
  magnetization in monolayer ${\mathrm{mos}}_{2}$},\ }\href
  {https://doi.org/10.1103/PhysRevLett.123.036806} {\bibfield  {journal}
  {\bibinfo  {journal} {Phys. Rev. Lett.}\ }\textbf {\bibinfo {volume} {123}},\
  \bibinfo {pages} {036806} (\bibinfo {year} {2019})}\BibitemShut {NoStop}%
\bibitem [{\citenamefont {Lee}\ \emph {et~al.}(2017)\citenamefont {Lee},
  \citenamefont {Wang}, \citenamefont {Xie}, \citenamefont {Mak},\ and\
  \citenamefont {Shan}}]{lee2017valley}%
  \BibitemOpen
  \bibfield  {author} {\bibinfo {author} {\bibfnamefont {J.}~\bibnamefont
  {Lee}}, \bibinfo {author} {\bibfnamefont {Z.}~\bibnamefont {Wang}}, \bibinfo
  {author} {\bibfnamefont {H.}~\bibnamefont {Xie}}, \bibinfo {author}
  {\bibfnamefont {K.~F.}\ \bibnamefont {Mak}},\ and\ \bibinfo {author}
  {\bibfnamefont {J.}~\bibnamefont {Shan}},\ }\bibfield  {title} {\bibinfo
  {title} {Valley magnetoelectricity in single-layer mos2},\ }\href
  {https://www.nature.com/articles/nmat4931#citeas} {\bibfield  {journal}
  {\bibinfo  {journal} {Nature materials}\ }\textbf {\bibinfo {volume} {16}},\
  \bibinfo {pages} {887} (\bibinfo {year} {2017})}\BibitemShut {NoStop}%
\bibitem [{\citenamefont {Radisavljevic}\ and\ \citenamefont
  {Kis}(2013)}]{radisavljevic2013mobility}%
  \BibitemOpen
  \bibfield  {author} {\bibinfo {author} {\bibfnamefont {B.}~\bibnamefont
  {Radisavljevic}}\ and\ \bibinfo {author} {\bibfnamefont {A.}~\bibnamefont
  {Kis}},\ }\bibfield  {title} {\bibinfo {title} {Mobility engineering and a
  metal--insulator transition in monolayer mos2},\ }\href
  {https://www.nature.com/articles/nmat3687} {\bibfield  {journal} {\bibinfo
  {journal} {Nature materials}\ }\textbf {\bibinfo {volume} {12}},\ \bibinfo
  {pages} {815} (\bibinfo {year} {2013})}\BibitemShut {NoStop}%
\bibitem [{\citenamefont {Ma}\ \emph {et~al.}(2019)\citenamefont {Ma},
  \citenamefont {Deng}, \citenamefont {Zheng}, \citenamefont {Wu},
  \citenamefont {Liu}, \citenamefont {Zhou},\ and\ \citenamefont
  {Sun}}]{ma2019experimental}%
  \BibitemOpen
  \bibfield  {author} {\bibinfo {author} {\bibfnamefont {J.}~\bibnamefont
  {Ma}}, \bibinfo {author} {\bibfnamefont {K.}~\bibnamefont {Deng}}, \bibinfo
  {author} {\bibfnamefont {L.}~\bibnamefont {Zheng}}, \bibinfo {author}
  {\bibfnamefont {S.}~\bibnamefont {Wu}}, \bibinfo {author} {\bibfnamefont
  {Z.}~\bibnamefont {Liu}}, \bibinfo {author} {\bibfnamefont {S.}~\bibnamefont
  {Zhou}},\ and\ \bibinfo {author} {\bibfnamefont {D.}~\bibnamefont {Sun}},\
  }\bibfield  {title} {\bibinfo {title} {Experimental progress on layered
  topological semimetals},\ }\href
  {https://iopscience.iop.org/article/10.1088/2053-1583/ab0902} {\bibfield
  {journal} {\bibinfo  {journal} {2D Materials}\ }\textbf {\bibinfo {volume}
  {6}},\ \bibinfo {pages} {032001} (\bibinfo {year} {2019})}\BibitemShut
  {NoStop}%
\bibitem [{\citenamefont {Polshyn}\ \emph {et~al.}(2019)\citenamefont
  {Polshyn}, \citenamefont {Yankowitz}, \citenamefont {Chen}, \citenamefont
  {Zhang}, \citenamefont {Watanabe}, \citenamefont {Taniguchi}, \citenamefont
  {Dean},\ and\ \citenamefont {Young}}]{polshyn2019large}%
  \BibitemOpen
  \bibfield  {author} {\bibinfo {author} {\bibfnamefont {H.}~\bibnamefont
  {Polshyn}}, \bibinfo {author} {\bibfnamefont {M.}~\bibnamefont {Yankowitz}},
  \bibinfo {author} {\bibfnamefont {S.}~\bibnamefont {Chen}}, \bibinfo {author}
  {\bibfnamefont {Y.}~\bibnamefont {Zhang}}, \bibinfo {author} {\bibfnamefont
  {K.}~\bibnamefont {Watanabe}}, \bibinfo {author} {\bibfnamefont
  {T.}~\bibnamefont {Taniguchi}}, \bibinfo {author} {\bibfnamefont {C.~R.}\
  \bibnamefont {Dean}},\ and\ \bibinfo {author} {\bibfnamefont {A.~F.}\
  \bibnamefont {Young}},\ }\bibfield  {title} {\bibinfo {title} {Large
  linear-in-temperature resistivity in twisted bilayer graphene},\ }\href
  {https://www.nature.com/articles/s41567-019-0596-3} {\bibfield  {journal}
  {\bibinfo  {journal} {Nature Physics}\ }\textbf {\bibinfo {volume} {15}},\
  \bibinfo {pages} {1011} (\bibinfo {year} {2019})}\BibitemShut {NoStop}%
\bibitem [{\citenamefont {Lu}\ \emph {et~al.}(2019)\citenamefont {Lu},
  \citenamefont {Stepanov}, \citenamefont {Yang}, \citenamefont {Xie},
  \citenamefont {Aamir}, \citenamefont {Das}, \citenamefont {Urgell},
  \citenamefont {Watanabe}, \citenamefont {Taniguchi}, \citenamefont {Zhang}
  \emph {et~al.}}]{lu2019superconductors}%
  \BibitemOpen
  \bibfield  {author} {\bibinfo {author} {\bibfnamefont {X.}~\bibnamefont
  {Lu}}, \bibinfo {author} {\bibfnamefont {P.}~\bibnamefont {Stepanov}},
  \bibinfo {author} {\bibfnamefont {W.}~\bibnamefont {Yang}}, \bibinfo {author}
  {\bibfnamefont {M.}~\bibnamefont {Xie}}, \bibinfo {author} {\bibfnamefont
  {M.~A.}\ \bibnamefont {Aamir}}, \bibinfo {author} {\bibfnamefont
  {I.}~\bibnamefont {Das}}, \bibinfo {author} {\bibfnamefont {C.}~\bibnamefont
  {Urgell}}, \bibinfo {author} {\bibfnamefont {K.}~\bibnamefont {Watanabe}},
  \bibinfo {author} {\bibfnamefont {T.}~\bibnamefont {Taniguchi}}, \bibinfo
  {author} {\bibfnamefont {G.}~\bibnamefont {Zhang}}, \emph {et~al.},\
  }\bibfield  {title} {\bibinfo {title} {Superconductors, orbital magnets and
  correlated states in magic-angle bilayer graphene},\ }\href
  {https://www.nature.com/articles/s41586-019-1695-0} {\bibfield  {journal}
  {\bibinfo  {journal} {Nature}\ }\textbf {\bibinfo {volume} {574}},\ \bibinfo
  {pages} {653} (\bibinfo {year} {2019})}\BibitemShut {NoStop}%
\bibitem [{\citenamefont {Kopnin}(2001)}]{kopnin2001nonequilibrium}%
  \BibitemOpen
  \bibfield  {author} {\bibinfo {author} {\bibfnamefont {N.}~\bibnamefont
  {Kopnin}},\ }\href@noop {} {\emph {\bibinfo {title} {Theory of Nonequilibrium
  Superconductivity}}}\ (\bibinfo  {publisher} {Oxford University Press},\
  \bibinfo {year} {2001})\BibitemShut {NoStop}%
\bibitem [{\citenamefont {Ferrell}\ and\ \citenamefont
  {Glover}(1958)}]{PhysRev.109.1398}%
  \BibitemOpen
  \bibfield  {author} {\bibinfo {author} {\bibfnamefont {R.~A.}\ \bibnamefont
  {Ferrell}}\ and\ \bibinfo {author} {\bibfnamefont {R.~E.}\ \bibnamefont
  {Glover}},\ }\bibfield  {title} {\bibinfo {title} {Conductivity of
  superconducting films: A sum rule},\ }\href
  {https://doi.org/10.1103/PhysRev.109.1398} {\bibfield  {journal} {\bibinfo
  {journal} {Phys. Rev.}\ }\textbf {\bibinfo {volume} {109}},\ \bibinfo {pages}
  {1398} (\bibinfo {year} {1958})}\BibitemShut {NoStop}%
\bibitem [{\citenamefont {Tinkham}\ and\ \citenamefont
  {Ferrell}(1959)}]{PhysRevLett.2.331}%
  \BibitemOpen
  \bibfield  {author} {\bibinfo {author} {\bibfnamefont {M.}~\bibnamefont
  {Tinkham}}\ and\ \bibinfo {author} {\bibfnamefont {R.~A.}\ \bibnamefont
  {Ferrell}},\ }\bibfield  {title} {\bibinfo {title} {Determination of the
  superconducting skin depth from the energy gap and sum rule},\ }\href
  {https://doi.org/10.1103/PhysRevLett.2.331} {\bibfield  {journal} {\bibinfo
  {journal} {Phys. Rev. Lett.}\ }\textbf {\bibinfo {volume} {2}},\ \bibinfo
  {pages} {331} (\bibinfo {year} {1959})}\BibitemShut {NoStop}%
\bibitem [{\citenamefont {Xia}\ \emph {et~al.}(2006)\citenamefont {Xia},
  \citenamefont {Maeno}, \citenamefont {Beyersdorf}, \citenamefont {Fejer},\
  and\ \citenamefont {Kapitulnik}}]{PhysRevLett.97.167002}%
  \BibitemOpen
  \bibfield  {author} {\bibinfo {author} {\bibfnamefont {J.}~\bibnamefont
  {Xia}}, \bibinfo {author} {\bibfnamefont {Y.}~\bibnamefont {Maeno}}, \bibinfo
  {author} {\bibfnamefont {P.~T.}\ \bibnamefont {Beyersdorf}}, \bibinfo
  {author} {\bibfnamefont {M.~M.}\ \bibnamefont {Fejer}},\ and\ \bibinfo
  {author} {\bibfnamefont {A.}~\bibnamefont {Kapitulnik}},\ }\bibfield  {title}
  {\bibinfo {title} {High resolution polar kerr effect measurements of
  ${\mathrm{sr}}_{2}{\mathrm{ruo}}_{4}$: Evidence for broken time-reversal
  symmetry in the superconducting state},\ }\href
  {https://doi.org/10.1103/PhysRevLett.97.167002} {\bibfield  {journal}
  {\bibinfo  {journal} {Phys. Rev. Lett.}\ }\textbf {\bibinfo {volume} {97}},\
  \bibinfo {pages} {167002} (\bibinfo {year} {2006})}\BibitemShut {NoStop}%
\bibitem [{\citenamefont {Kapitulnik}\ \emph {et~al.}(2009)\citenamefont
  {Kapitulnik}, \citenamefont {Xia}, \citenamefont {Schemm},\ and\
  \citenamefont {Palevski}}]{kapitulnik2009polar}%
  \BibitemOpen
  \bibfield  {author} {\bibinfo {author} {\bibfnamefont {A.}~\bibnamefont
  {Kapitulnik}}, \bibinfo {author} {\bibfnamefont {J.}~\bibnamefont {Xia}},
  \bibinfo {author} {\bibfnamefont {E.}~\bibnamefont {Schemm}},\ and\ \bibinfo
  {author} {\bibfnamefont {A.}~\bibnamefont {Palevski}},\ }\bibfield  {title}
  {\bibinfo {title} {Polar kerr effect as probe for time-reversal symmetry
  breaking in unconventional superconductors},\ }\href
  {https://iopscience.iop.org/article/10.1088/1367-2630/11/5/055060} {\bibfield
   {journal} {\bibinfo  {journal} {New Journal of Physics}\ }\textbf {\bibinfo
  {volume} {11}},\ \bibinfo {pages} {055060} (\bibinfo {year}
  {2009})}\BibitemShut {NoStop}%
\bibitem [{\citenamefont {Tagay}\ \emph {et~al.}(2023)\citenamefont {Tagay},
  \citenamefont {Romero~III},\ and\ \citenamefont {Armitage}}]{tagay2023high}%
  \BibitemOpen
  \bibfield  {author} {\bibinfo {author} {\bibfnamefont {Z.}~\bibnamefont
  {Tagay}}, \bibinfo {author} {\bibfnamefont {R.}~\bibnamefont {Romero~III}},\
  and\ \bibinfo {author} {\bibfnamefont {N.}~\bibnamefont {Armitage}},\
  }\bibfield  {title} {\bibinfo {title} {High-precision measurements of
  terahertz polarization states with a fiber coupled time-domain thz
  spectrometer},\ }\href {https://arxiv.org/abs/2312.13276} {\bibfield
  {journal} {\bibinfo  {journal} {arXiv preprint arXiv:2312.13276}\ } (\bibinfo
  {year} {2023})}\BibitemShut {NoStop}%
\bibitem [{\citenamefont {Matsunaga}\ \emph {et~al.}(2013)\citenamefont
  {Matsunaga}, \citenamefont {Hamada}, \citenamefont {Makise}, \citenamefont
  {Uzawa}, \citenamefont {Terai}, \citenamefont {Wang},\ and\ \citenamefont
  {Shimano}}]{PhysRevLett.111.057002}%
  \BibitemOpen
  \bibfield  {author} {\bibinfo {author} {\bibfnamefont {R.}~\bibnamefont
  {Matsunaga}}, \bibinfo {author} {\bibfnamefont {Y.~I.}\ \bibnamefont
  {Hamada}}, \bibinfo {author} {\bibfnamefont {K.}~\bibnamefont {Makise}},
  \bibinfo {author} {\bibfnamefont {Y.}~\bibnamefont {Uzawa}}, \bibinfo
  {author} {\bibfnamefont {H.}~\bibnamefont {Terai}}, \bibinfo {author}
  {\bibfnamefont {Z.}~\bibnamefont {Wang}},\ and\ \bibinfo {author}
  {\bibfnamefont {R.}~\bibnamefont {Shimano}},\ }\bibfield  {title} {\bibinfo
  {title} {Higgs amplitude mode in the bcs superconductors
  ${\mathrm{nb}}_{1\mathrm{\text{\ensuremath{-}}}x}{\mathrm{ti}}_{x}\mathbf{N}$
  induced by terahertz pulse excitation},\ }\href
  {https://doi.org/10.1103/PhysRevLett.111.057002} {\bibfield  {journal}
  {\bibinfo  {journal} {Phys. Rev. Lett.}\ }\textbf {\bibinfo {volume} {111}},\
  \bibinfo {pages} {057002} (\bibinfo {year} {2013})}\BibitemShut {NoStop}%
\bibitem [{\citenamefont {Raab}\ and\ \citenamefont
  {de~Lange}(2004)}]{raab2004multipole}%
  \BibitemOpen
  \bibfield  {author} {\bibinfo {author} {\bibfnamefont {R.~E.}\ \bibnamefont
  {Raab}}\ and\ \bibinfo {author} {\bibfnamefont {O.~L.}\ \bibnamefont
  {de~Lange}},\ }\href@noop {} {\emph {\bibinfo {title} {Multipole Theory in
  Electromagnetism: Classical, Quantum, and Symmetry Aspects, with
  Applications}}}\ (\bibinfo  {publisher} {Oxford University Press},\ \bibinfo
  {year} {2004})\BibitemShut {NoStop}%
\bibitem [{\citenamefont {Smidman}\ \emph {et~al.}(2017)\citenamefont
  {Smidman}, \citenamefont {Salamon}, \citenamefont {Yuan},\ and\ \citenamefont
  {Agterberg}}]{smidman2017superconductivity}%
  \BibitemOpen
  \bibfield  {author} {\bibinfo {author} {\bibfnamefont {M.}~\bibnamefont
  {Smidman}}, \bibinfo {author} {\bibfnamefont {M.}~\bibnamefont {Salamon}},
  \bibinfo {author} {\bibfnamefont {H.}~\bibnamefont {Yuan}},\ and\ \bibinfo
  {author} {\bibfnamefont {D.}~\bibnamefont {Agterberg}},\ }\bibfield  {title}
  {\bibinfo {title} {Superconductivity and spin--orbit coupling in
  non-centrosymmetric materials: a review},\ }\href
  {https://iopscience.iop.org/article/10.1088/1361-6633/80/3/036501/meta}
  {\bibfield  {journal} {\bibinfo  {journal} {Reports on Progress in Physics}\
  }\textbf {\bibinfo {volume} {80}},\ \bibinfo {pages} {036501} (\bibinfo
  {year} {2017})}\BibitemShut {NoStop}%
\end{thebibliography}%

\end{document}